\documentclass[a4paper,12pt]{article}
\usepackage{jheppub} % for details on the use of the package, please
\usepackage{float}
\usepackage{ulem}
\usepackage{bm}
\usepackage{cancel}
\usepackage{enumitem}

\usepackage[latin1]{inputenc}

\usepackage{comment}

\usepackage{color}
\usepackage{graphicx}

\usepackage{shuffle}

\renewcommand{\thefootnote}{\fnsymbol{footnote}}
\renewcommand{\thanks}[1]{\footnote{#1}}

\newcommand{\bea}{\begin{eqnarray}}
\newcommand{\eea}{\end{eqnarray}}
\newcommand{\ee}{\end{equation}}
\newcommand{\be}{\begin{equation}}

\newcommand{\no}{\nonumber}
\def\eqn#1{eq.~(\ref{#1})}

\def\cN{{\cal N}}

\def\det{{\rm det}}
\def\Det{{\rm Det}}

\def\ep{\varepsilon}

\def\bep{\boldsymbol{\mathbf{\varepsilon}}}

\newcommand\nn{\nonumber}

\newcommand{\Gmat}{\gamma}
\newcommand{\auxu}{z}
\newcommand{\auxv}{\tilde z}

\newcommand\spaq[1]{\langle #1\rangle}
\newcommand\spbq[1]{[#1]}

\def\spa#1.#2{\left\langle#1\,#2\right\rangle}
\def\spb#1.#2{\left[#1\,#2\right]}
\def\spab#1.#2{\left\langle#1\,#2\right]}
\def\spba#1.#2{\left[#1\,#2\right\rangle}
\def\spaq#1.#2{\langle#1 | #2\rangle}
\def\spbq#1.#2{[#1 | #2]}
\def\eqn#1{Eq.~(\ref{#1})}

\def\bep{\boldsymbol{\mathbf{\ep}}}
\def\ep{\varepsilon}

\def\etaT{\tilde \eta}

\def\aket#1{\underline{\vphantom{q_3}\bm{#1}}\rangle}
\def\sket#1{\underline{\vphantom{q_3}\bm{#1}}]}
\def\ket#1{\underline{\vphantom{q_3}{#1}}\rangle}
\def\abra#1{\langle\underline{\vphantom{q_3}\bm{#1}}}
\def\sbra#1{[\underline{\vphantom{q_3}\bm{#1}}}
\def\bra#1{\langle\underline{\vphantom{q_3}{#1}}}
\def\qref#1{q_{#1}}
\def\qi{Q}

\title{\boldmath Spinor-helicity formalism for massive and massless amplitudes in five dimensions}

\author[a]{Marco Chiodaroli,}
\author[b]{Murat G\"{u}naydin,}
\author[a,c]{Henrik Johansson,}
\author[b]{and Radu Roiban}

\affiliation[a]{Department of Physics and Astronomy, \\ Uppsala University, 75108 Uppsala, Sweden}
\affiliation[b]{Institute for Gravitation and the Cosmos, \\ The Pennsylvania State University,
University Park, PA 16802, USA}
\affiliation[c]{Nordita, Stockholm University and KTH Royal Institute of Technology,\\  Hannes Alfv\'{e}ns  v\"{a}g 12, 10691 Stockholm, Sweden}

\emailAdd{marco.chiodaroli@physics.uu.se}
\emailAdd{mgunaydin@psu.edu}
\emailAdd{henrik.johansson@physics.uu.se}
\emailAdd{radu@phys.psu.edu}

\abstract{
Five-dimensional gauge and gravity theories are known to exhibit striking properties. $D=5$ is the lowest dimension where massive tensor states appear naturally, providing a testing ground for 
perturbative insights into six-dimensional tensor theories. Five-dimensional supergravities are highly constrained and admit elegant geometric and algebraic formulations, with global symmetries manifest at the Lagrangian level. 

In this paper, we take a step towards the systematic investigation of amplitudes in five dimensions, and  present a five-dimensional version of the  spinor-helicity formalism, applicable to massless, massive and supersymmetric states. We give explicit representations for on-shell spinor and polarization variables such that the little-group symmetry and gauge redundancy are manifest. Massive self-dual tensor states are discussed in some detail, as well as all the on-shell supermultiplets that can appear in matter-coupled gauge and supergravity theories. 
As a byproduct of considering supersymmetry in the presence of central charge, we obtain massless ten-dimensional Majorana-Weyl spinors as products of five-dimensional massive spinors. 

We present compact expressions for superamplitudes at multiplicity three and four, including several novel superamplitudes that either do not straightforwardly uplift to six dimensions, or have not appeared in the six-dimensional literature.
We discuss several examples of five-dimensional double-copy constructions in the context of gravitational theories with massive vectors and tensors, illustrating that the formalism we construct can also be used to considerably streamline the
double-copy construction of $\cN=2$ Maxwell-Einstein supergravities.

}

\preprint{UUITP-08/22 \\
\phantom{~} \hfill NORDITA-2021-165}

\begin{document}
\maketitle
\flushbottom

\renewcommand{\thefootnote}{\arabic{footnote}}

\section{Introduction}

Over the past decades, four-dimensional spinor-helicity variables have been 
crucial for streamlining scattering amplitude calculations, uncovering novel structures in gauge theories and gravity. 
They are a fundamental ingredient of the Parke-Taylor formula for tree-level MHV amplitudes~\cite{Parke:1986gb}, are closely related to the twistor-space description of amplitudes \cite{Witten:2003nn,Roiban:2004yf,Cachazo:2004kj,Gukov:2004ei}, and provide a simple presentation for modern on-shell recursion relations \cite{Cachazo:2004kj,Britto:2005fq}.
The built-in advantage of the spinor-helicity formalism consists in providing a covariant and dimension-specific solution for the on-shell conditions, leading to simple expressions for momenta and physical polarization vectors and tensors. In turn, this leads to the observation that scattering amplitudes become simple once they are written 
in terms of physical (i.e.~on-shell) quantities.

Spinor-helicity methods have been used most successfully for amplitudes involving four-dimensional massless states.  Detailed extensions of the formalism have been introduced in three, six and ten dimensions \cite{Dennen:2009vk,Cheung:2009dc,Bern:2010qa,Dennen:2010dh,Caron-Huot:2010nes}, as well as for massive fields in four dimensions~\cite{Arkani-Hamed:2017jhn} (see also Refs.~\cite{Craig:2011ws,Kiermaier:2011cr,Ochirov:2018uyq,Herderschee:2019dmc,Johansson:2019dnu,Chiodaroli:2021eug}). While helicity cannot be defined in higher dimensions and for massive fields, the term higher-dimensional spinor-helicity is commonly used (with a slight abuse of notation) to refer to the higher-dimensional extension of the 4D spinor-helicity formalism. The invariant content of the higher-dimensional approach is to properly classify all variables by both their little-group and Lorentz group representations, and as such spinor/little-group formalism can be used as a synonym.  

Five-dimensional theories are interesting because they populate a special corner of the parameter space of known theories. 
On the one hand, all theories in five dimensions can be dimensionally reduced to four dimensions. This reduction, however, obscures properties that are indigenous to five dimensions, such as supergravity U-duality symmetries being symmetries of the action rather than of the equations of motion.
On the other hand, not all five-dimensional theories can be lifted to six dimensions, either because of their field content, or because of their interactions. Thus, studying effects that depend strongly on the dimensional regulator, such as rational terms in the amplitudes, may be subtle. It may be more appropriate to make all variables explicitly five dimensional. 
In this work, we tackle the problem of defining natural variables for scattering amplitudes in five spacetime dimensions, and initiate the study of the 5D spinor-helicity formalism. 

Large classes of $\cN=2$ Maxwell-Einstein and Yang-Mills-Einstein supergravities are naturally formulated in five dimensions~\cite{Gunaydin:1983bi,Gunaydin:1984ak,Gunaydin:1986fg,Gunaydin:1999zx,Gunaydin:2000xk,Gunaydin:2005bf,Gunaydin:2003yx,deWit:1991nm}.
Their Lagrangians are considerably simpler than those of their four-dimensional relatives, and yet present a far richer structure than theories that uplift to higher dimension. While the study of scattering amplitudes in these theories has been yielding interesting results \cite{Carrasco:2012ca,Chiodaroli:2014xia,Chiodaroli:2015rdg,Chiodaroli:2015wal,Chiodaroli:2016jqw,Anastasiou:2016csv,Chiodaroli:2017ngp,Ben-Shahar:2018uie,Chiodaroli:2017ehv,Anastasiou:2017nsz} (see also Ref.~\cite{Bern:2019prr} for a comprehensive review), until now the analysis relied on using four-dimensional spinor-helicity methods after dimensional reduction. To streamline and advance the study of  amplitudes in five-dimensional theories, it is therefore desirable to develop tools that are designed to work directly in five dimensions, bypassing the need for dimensional reduction. 
Additionally,  various gaugings of maximal 
supergravity have been explicitly studied in five dimensions
\cite{Gunaydin:1984qu,Gunaydin:1985cu,Gunaydin:1985tb,Hull:1988jw}, including some very recent results~\cite{Dallagata:2021lsc,Bobev:2020ttg}. While the application of amplitude methods to the study of gauged maximal supergravity is still in its infancy~\cite{Chiodaroli:2018dbu}, the five-dimensional spinor-helicity methods developed here should provide a valuable tool for advanced calculations~\cite{Johansson:2017bfl}.     

A central aspect of $D=5$ is that it is the lowest dimension where massive anti-symmetric tensor states and fields can arise, which cannot be dualized to other types of interacting fields \cite{Gunaydin:1999zx,Gunaydin:2000xk,Gunaydin:2005bf,Gunaydin:2003yx}. Studying theories with massive tensors in five dimensions may prove to be  crucial for understanding six-dimensional theories with tensors, possibly including the famed $(2,0)$ theory. The study of amplitudes in the $(2,0)$ theory has so far presented notable difficulties, for example, due to the lack of suitable three-point amplitudes that can be used as building blocks~\cite{Czech:2011dk} (see also Refs. \cite{Heydeman:2017yww,Chang:2018xmx}).\footnote{For a different approach for understanding perturbative six-dimensional self-dual tensor theories see Refs.~\cite{Ho:2011ni, Ho:2012nt, Ho:2014eoa, Huang:2012tu,Samtleben:2011fj, Samtleben:2011drz,Lambert:2010wm,Chu:2012um,Bonetti:2012st}.} 
In many cases, the study of amplitudes in five dimensions has required adapting tools and methods developed in six dimensions. This includes dimensional reduction of the  spinor-helicity formalism from six dimensions \cite{Czech:2011dk}, techniques based on scattering equations \cite{Cachazo:2018hqa,Geyer:2018xgb}, as well as  ambitwistor strings \cite{Albonico:2020mge,Geyer:2020iwz}.

 Venturing away from Minkowski backgrounds, five-dimensional theories are of great interest from the perspective of the AdS/CFT correspondence \cite{Maldacena:1997re,Witten:1998qj,Gubser:1998bc}. In its simplest formulation it relates correlation functions in ${\cal N}=4$ super-Yang-Mills theory to boundary correlation functions in 5D gauged supergravity given by the reduction of type IIB supergravity on a five-sphere~\cite{Gunaydin:1984fk,Kim:1985ez}. 
Supergravity fields are then naturally organized with a split notation in which spacetime indices along the compact directions are related to  R-symmetry indices. The massive 5D spinors with central charge that we will study have a similar organization, and it is likely that they can be suitably adapted to describe states in AdS${}_5 \times S_5$ background.

This paper is structured as follows. In Section~\ref{sec2}, we provide the basis of the 5D formalism we are using and introduce the relevant notation. In Section~\ref{sec3}, we discuss supersymmetry and the organization of the on-shell states into 5D superfields. We also briefly outline the relation between our 5D superfields and 4D and 6D superfields. In Section~\ref{sec4}, we discuss superamplitudes at three and four points for interacting vector multiplets. Finally, in Section~\ref{sec5}, we discuss theories of gravitationally interacting fields, including massive tensors, from the point of view of the double-copy construction~\cite{Kawai:1985xq,Bern:2008qj,Bern:2010ue}. Among other things, we obtain compact expressions for amplitudes for Maxwell-Einstein and Yang-Mills-Einstein theories, recovering previous results in a considerably simpler form. We then  conclude with a discussion of open problems and direction for future investigation.

\section{Five-dimensional spinor-helicity formalism \label{sec2}}

To make the 5D spinor-helicity notation accessible, in this section we spell out the necessary details, including the explicit parametrization for the frequently-encountered spinors and polarization vectors and tensors. 

\subsection{Five-dimensional on-shell spinors: massless and massive}

Working in mostly-minus signature $(+----)$, we consider the $SO(1,4)\rightarrow SO(1,3)$ decomposition of a 5D momentum and gamma matrices, $p_\mu=( p_{\bar \mu}, p_4)$, $\gamma^\mu=(\gamma^{\bar \mu},\gamma^4)$, respectively. Here 4D indices are barred for clarity, $\bar \mu=0,{\ldots},3$, and what we call $\gamma^4$ is the usual $\gamma^5$ matrix in 4D. Thus, in the Weyl basis, we have
\be
\label{pslash}
\slash \hskip-2.4mm p  \equiv p_\mu \gamma^\mu = \left(\begin{matrix}
-i p_4 ~~ &   p \cdot \sigma \\
p \cdot \bar \sigma ~~&~~ i p_4   \\
\end{matrix}\right)\,,
\ee
where $\sigma^{\bar \mu}=(1,\sigma_i)$ and $\bar \sigma^{\bar \mu}=(1,-\sigma_i)$ are the 4D sigma matrices. Because of the  isomorphism $SO(1,4)\cong USp(2,2)$, we can lower the upper index on $\slash \hskip-2.4mm p$ and write the 5D momentum as an antisymmetric $4\times 4$ matrix, 
\be
p_{AB}=p_\mu \Gmat_{AB}^\mu \equiv -p_\mu (\Gmat^\mu)^{\ C}_A \Omega_{CB} =
-(\slash \hskip-2.4mm p \, \Omega)_{AB}\ .
\label{pLowerAB}
\ee
We choose the symplectic metric $\Omega$ of $USp(2,2)$ to be block diagonal,
\be \label{OmegaDef}
\Omega_{AB}= \left(\begin{matrix}
\epsilon_{\alpha \beta} & 0 \\
0&  - \epsilon^{\dot \alpha \dot \beta}   \\
\end{matrix}\right)= \Omega^{BA}\,,
\ee
where $A,B$ are fundamental $USp(2,2)$ indices decomposed as $A=\alpha\oplus \dot\alpha$ in terms of the $SL(2,\mathbb{C})$ indices of the Levi-Civita symbols and sigma matrices $\sigma^{\bar \mu}_{\alpha \dot \beta}$,  $\bar \sigma^{\dot \beta \alpha}_{\bar \mu}$.
We use $\Omega$ to lower and raise indices, according to the left-multiplication convention, {e.g.} 
\be
X^A=\Omega^{AB} X_B \ , \qquad \qquad  X_A=\Omega_{AB} X^B \ .
\label{raiselowerrule}
\ee
The two-dimensional Levi-Civita symbol is normalized as $ \epsilon^{12}=\epsilon_{21}=1$, implying that $\Omega_{AC}\Omega^{CB}=\delta_{A}^{\ B}$. 

The gamma matrices with lowered indices, $\Gmat^\mu_{AB}= \Omega_{BC} (\Gmat^\mu)_A^{\ C}$, are antisymmetric and $\Omega$-traceless, $\Gmat^\mu_{AB} \Omega^{BA}=0$. Furthermore, in addition to the Clifford algebra $\{ \Gmat^\mu , \Gmat^\nu\} = 2 \eta^{\mu\nu}$, they obey the special 5D identity
\be
 (\Gmat^\mu)_A^{\ B} (\Gmat_\mu)_C^{\ D} =- 2 \Omega_{AC} \Omega^{BD}+  2\delta_{A}^D \delta_{C}^B - \delta_{A}^B \delta_{C}^D\,.
\ee
See Appendix~\ref{AppendixA} for further details on the gamma matrices. 

The determinant of $p_{AB}$ evaluates to  $\Det(p_{AB})= (\frac{1}{4}p_{AB} p^{AB})^2 =(p^2)^2$. Thus, for massless 5D momentum, $p^2=0$, the matrix $p_{AB}$ has rank two.
It can therefore be factorized over the $SU(2)$ little group using on-shell $USp(2,2)$ spinors,
\be
p_{AB}=|p_a\rangle_{A} | p^a \rangle_{B}
~~~~~~~{\rm or}~~~~~~
{\slash \hskip-2.4mm p} {}_A{}^B\equiv p_{A}{}^{B}=|p_a\rangle_{A} \langle p^a |^{B} \ ,
\ee
where $a,b,\ldots$ are $SU(2)$ little-group indices
and the ``bra'' is defined as   $\langle p^a  | {}^B=  \Omega^{BA} | p^a \rangle_{A} $.
Little-group indices can be lowered and raised through left-multiplication, similarly to Eq. \eqref{raiselowerrule}, by the Levi-Civita symbol $\epsilon^{ab}=\epsilon_{ba}$. An explicit parametrization of the massless $USp(2,2)$ spinor is
\be \label{USp4spinors}
|p^a\rangle_{A} =\left(\begin{matrix}
p_0+p_3 ~&~ 0 \\
p_1+ip_2 ~&~ -\frac{i p_4}{p_0+ p_3 } \\
i p_4 ~&~ -\frac{p_1-i p_2}{p_0+ p_3} \\
0 ~&~  1 \\
\end{matrix}\right)\, ,
\ee
where we have scaled the 
little-group 
components to make the spinor entries free of square roots. Using spinors that are rational functions of the momentum components will greatly simplify practical calculations.
The parametrization \eqref{USp4spinors} is such that the contraction of two 
$USp(2,2)$ spinors of the same massless momentum gives a vanishing result,
\be \label{OSconstraint}
\langle p^a  | p^b\rangle \equiv  \Omega^{BA} | p^a \rangle_{A}   | p^b \rangle_{B} =0\, .
\ee
Eq. (\ref{OSconstraint}) implies that the spinor satisfies the massless Dirac equation,
\be
p^{AB} | p_a \rangle_{B} =0\,,
\ee
 and hence it is an on-shell spinor.\footnote{Here we treat the on-shell spinor as a function of the momentum following the parametrization in Eq.~(\ref{USp4spinors}).  However, if the spinor is used to define the null momentum, then \eqn{OSconstraint} becomes a one-parameter constraint.  The presence of this constraint differentiates the 5D spinor-helicity formalism from its 4D and 6D relatives.} Here $p^{AB}$ is obtained by raising the indices in Eqs.~\eqref{pslash}-\eqref{pLowerAB}.

We proceed to extend the construction above to massive 5D momenta obeying $p^2=m^2$. The matrix $p_{AB}$ now has rank 4, so it can be factorized over the massive little group $SO(4)\sim SU(2)\times SU(2)$. 
To construct this factorization, we note that we can always split $p$ into two massless momenta $k$ and $q$,
\be \label{massiveMom}
p^\mu = k^\mu+ m^2 q^\mu\,,
\ee
where $q^\mu$ is a reference null-vector that satisfies $2 p \cdot q =2 k \cdot q=1$, and $k^\mu$ is defined by the above relation. Since both $k$ and $q$ are massless, we can reuse for each of them the parametrization from Eq. (\ref{USp4spinors}). After contracting \eqn{massiveMom} with the gamma matrices, we get an expression that can be factorized over the $SU(2)\times SU(2)$ little group,
\be
 p_{AB} = |k_a \rangle_A  |k^a \rangle_B +  m^2 |q_a \rangle_A  |q^a\rangle_B  = {1 \over 2} 
 | {\bm p}_a \rangle_A    | {\bm p}^a \rangle_B  + 
 {1 \over 2} 
 |{\bm p}_{\dot a}]_A    |{\bm p}^{\dot a}]_B\,,
\ee
or, alternatively,
\be
 p_{A}^{\ B} = {1 \over 2} 
 | {\bm p}_a \rangle_A    \langle {\bm p}^a |^B  +
 {1 \over 2} 
 |{\bm p}_{\dot a}]_A    [{\bm p}^{\dot a}|^B\,,
\ee
where we have defined new massive spinors
\bea
|{\bm p}^a\rangle_A  &=& 
|k^a \rangle_A +  m  |q^a \rangle_A \,, \nn \\ 
|{\bm p}^{\dot a} ]_A  &=& |k^{\dot a} \rangle_A - m  |q^{\dot a} \rangle_A \,.
\label{massivespinors}
\eea
While the little-group indices of the massless spinors run over the diagonal $SU(2)$ subgroup, here we assign the indices $a$ and $\dot a$ to run over the left and right factors of the massive little group $SU(2)\times SU(2)$, respectively.

Let us further spell out the properties of the reference spinor $|q^a \rangle_A$. We will demand that it is normalized relative to the $|k^a \rangle_A$ spinor, to satisfy
\be \label{q_constraint}
\langle k^a  | q^b\rangle =  \epsilon^{ab} \,.
\ee 
Indeed, this relation can be solved as
\begin{equation}
\label{q_constraint_solution}
| q^a\rangle = | \tilde q^b\rangle  \langle \tilde q  |  k\rangle^{-1}_{bc}  \epsilon^{ac} \ , 
\end{equation}
where $| \tilde q^b\rangle $ is an arbitrary massless $USp(2,2)$ spinor, and $\langle \tilde q | k\rangle^{-1}_{bc}$ is the matrix inverse of $\langle k^c  | \tilde q^b\rangle$. It then follows that $q^\mu = \frac{1}{4}\langle q_a| \Gmat^\mu|q^a\rangle= \tilde q^\mu /(2  \tilde q \cdot p ) $. 
In particular, it implies that $2p\cdot q= 2k\cdot q=  \frac{1}{2}\langle k^a  | q^b\rangle  \langle q_b  | k_a\rangle =1$ as desired. The constraint (\ref{q_constraint}) is stronger than the momentum constraint $2k\cdot q=1$, as it also imposes an alignment of the little groups of the $k,q$ vectors. We will assume that this stronger constraint holds throughout the paper.

The constraint in \eqn{q_constraint} implies that the contractions of the two spinors corresponding to a massive momentum are
\be
\langle {\bm p}^a  | {\bm p}^b \rangle = 2 
m \epsilon^{ab}\,,~~~  \qquad 
[ {\bm p}^{\dot a} | {\bm p}^{\dot b} ] = - 2 
m \epsilon^{\dot a \dot b}  \,,~~~  \qquad 
\langle {\bm p}^{a} |  {\bm p}^{\dot b} ] =0\,.
\ee
It therefore follows that the massive spinors obey the massive Dirac equation,
\bea
p^{A B}|  {\bm p}^a \rangle_B &=& -m \langle {\bm p}^a|^A\,, \nn \\
p^{AB}|  {\bm p}^{\dot a}]_B &=& m \, [{\bm p}^{\dot a} |^A\, .
\eea
While it is tempting to identify the above spinors with the standard particle $u$ and antiparticle $v$ solutions to the Dirac equation, this is not quite accurate. As shown in the Appendix~\ref{AppendixA}, the reality properties of the square and angle spinors are compatible with them being symplectic-Majorana spinors. 

In terms of the pair of massless spinors that satisfy \eqn{q_constraint}, 
the $USp(2,2)$ identity operator can be written as
\be
| q_a\rangle_A  \langle k^a  |^B +| k_a\rangle_A   \langle q^a  |^B =\delta_{A}^B \ ,
\ee
from which the completeness relations of the massive spinors follow,
\bea \label{spinorCompl}
&& | {\bm p}_a\rangle_A  \langle {\bm p}^a  |^B = 
p_{A}^{\ B} + m\, \delta_{A}^B\equiv 2 m ({\cal P}^+)_{A}^{\ B}\,, \nn \\
&& | {\bm p}_{\dot a}]_A  [ {\bm p}^{\dot a}|^B = 
p_{A}^{\ B} - m\, \delta_{A}^B  \equiv 2 m ({\cal P}^-)_{A}^{\ B}\,.
\eea
The projectors satisfy ${\cal P}^{\pm} {\cal P}^{\pm}= \pm {\cal P}^{\pm}$ and ${\cal P}^{\pm} {\cal P}^{\mp}= 0$. The above relations are analogous to the completeness relations of the standard $u$ and $v$ spinors.

Before ending this section, we note that there exists a reference spinor that is particularly natural to work with.
For massless momenta $k_\mu$, a convenient parametrization of the reference spinor is
\be \label{choiceq}
|q^a\rangle_{A} =\left(\begin{matrix}
0 ~&~ 0 \\
0 ~&~ \frac{1}{k_0+ k_3 } \\
1 ~&~0 \\
0 ~&~  0 \\
\end{matrix}\right)\,.
\ee
which corresponds to the null vector $q_\mu=\frac{1}{2(k_0 + k_3)} (1, 0,0,-1,0)$, and the global null vector $\tilde q_\mu=(1,0,0,-1,0)$. However, this spinor cannot be obtained by plugging the momentum $q_\mu$ into the parametrization in Eq. (\ref{USp4spinors}), since this gives a singular expression. Indeed, the above reference spinor is located at infinity in the parametrization (\ref{USp4spinors}).
This choice has the benefit that inner products are simple,
\begin{equation}
\langle q_i^a |q_j^b \rangle = 0 \ , \qquad
\langle k_i^a |q_j^b \rangle =
\left(\begin{matrix}
0 ~&~ 1 \\
-\frac{k_{0i}+k_{3i}}{k_{0j}+k_{3j}}~&~ 0
\end{matrix}\right)
\end{equation}
where $i,j$ are the particle labels.

Furthermore, with the choice~(\ref{choiceq}), the massive-spinor parametrizations simplify considerably. From \eqn{massivespinors} and the identity $k_0+k_3=p_0+p_3$, we obtain the massive spinor
\be \label{USp4spinorMassive}
|{\bm p}^a\rangle_{A} =
\left(\begin{matrix}
p_0+p_3 ~&~ 0 \\
p_1+ip_2 ~&~ \frac{m-i p_4}{p_0+ p_3 } \\
m+i p_4 ~&~ -\frac{p_1-i p_2}{p_0+ p_3} \\
0 ~&~  1 \\
\end{matrix}\right)\,.
\ee
The spinor $|{\bm p}^{\dot a}]_{A}$ can be obtained by flipping the sign of the mass as $|{\bm p}]= |{\bm p} \rangle\big|_{m\rightarrow -m}$. We will use these massive spinors in the remainder of the paper. 

When displaying amplitudes in the following sections, we will find convenient to dress the free little-group indices with bosonic auxiliary variables $\auxu_a$ and $\auxv_{\dot a}$.  In formulae in which we do not wish to explicitly display these variables, we will adopt the following short-hand notation,
\be
|\,\ket{i\,} \equiv | k^a_i \rangle \auxu_{ia}  
\,,~~~~ \qquad
|\,\aket{i\,} \equiv | {\bm p}^a_i \rangle \auxu_{ia}
\, ,~~~~ \qquad 
|\,\sket{i\,} \equiv | {\bm p}^{\dot a}_i ] \auxv_{i{\dot a}} \ , \label{compactnotation}
\ee
where the index $i$ is the particle label.
However, for the reader's convenience, we will keep the little-group indices and momentum explicit for the remainder of this section.

\subsection{Massless and massive polarization vectors}

An advantage of spinor-helicity variables is that they provide convenient parametrizations for general asymptotic states. Here we construct 5D polarization vectors, using the previously-introduced spinor variables. 
We start with the polarizations of massless vectors. The only natural choice is\footnote{Upon dimensional reduction, these expressions reproduce the ones from Ref.~\cite{Dixon:1996wi} with the identifications  $\varepsilon^\mu_{11}\rightarrow \varepsilon^\mu_-$, $\varepsilon^\mu_{22}\rightarrow \varepsilon^\mu_+$.}
\be
\varepsilon^\mu_{a  b} (k,q) =  \frac{ \langle k_{(a}| \Gmat^\mu   | q_{b)} \rangle}{\sqrt{2}}=  -\frac{ \langle q_{(a}| \Gmat^\mu   | k_{b)} \rangle}{\sqrt{2}}\,.
\label{masslesspol}
\ee
where $a,b$ are the $SU(2)$ little-group indices and the symmetrization is normalized as $A_{(a}B_{b)}={1\over 2}A_aB_b+{1\over 2}A_bB_a$. The reference spinor $| q^{b} \rangle$ must satisfy Eq. (\ref{q_constraint}) or, equivalently, Eq. \eqref{q_constraint_solution}. 
The massless polarization vector constructed in this way is transverse, $k \cdot \varepsilon_{ab} (k,q) =q \cdot\varepsilon_{ab} (k,q) =0$, because $|k\rangle$ and $|q\rangle$ obey (by construction) the massless Dirac equation. It also satisfies the following completeness relations:
\bea
&&\varepsilon^\mu_{a  b} (k,q)  \varepsilon_{\mu ,cd} (k,q)  = -\frac{1}{2}(\epsilon_{ac}  \epsilon_{bd}+\epsilon_{ad}  \epsilon_{bc})\,, \nn \\ 
&&\varepsilon^\mu_{a  b} (k,q) \varepsilon^{\nu ,a b} (k,q) =  -\eta^{\mu  \nu } + 2 k^{(\mu } q^{\nu )} \ .
\eea
Any massless polarization can be mapped to one written in terms of the reference spinor given in Eq.~(\ref{choiceq}) 
with a gauge transformation.\footnote{Note that enacting different gauge transformations for the different polarizations does not change amplitudes and hence does not break the $SU(2)$ little-group symmetry for the asymptotic states.}

Using the massive on-shell spinors we have constructed in the previous section, we can also construct polarization vectors for massive vectors ($W$ bosons),
\be
\varepsilon^\mu_{a  \dot a} (p) =  -\frac{ \langle {\bm p}_{a}| \Gmat^\mu   | {\bm p}_{{\dot a}}]}{2\sqrt{2}m}\, .
\label{massivepol}
\ee
By construction, these are transverse as before, $p \cdot \varepsilon_{a  \dot a} (p)=0$, and  span the $SO(4)$ little group,
\be
\varepsilon^\mu_{a  \dot a} (p)  \varepsilon_{\mu ,b  \dot b} (p)  = - \epsilon_{ab} \epsilon_{ \dot a  \dot b} \ .
\ee
Moreover, they obey the Minkowski-space completeness relation appropriate to vectors transverse to a massive on-shell momentum,
\be
-\varepsilon^\mu_{a  \dot a} (p)  \varepsilon^{\nu ,a  \dot a} (p) = \eta^{\mu  \nu } - \frac{p^\mu  p^\nu }{m^2} \equiv {\tilde \eta}^{\mu\nu}\, ,
\label{massivephysproj}
\ee
which is also the massive-vector physical-state projector.
We may also construct a linearized field strength for the massive vector using the spinor. It takes the form
\be
F^{\mu\nu}_{a {\dot a}} = - \frac{ \langle {\bm p}_{a}| \Gmat^{\mu\nu}  | {\bm p}_{\dot a} ] }{2\sqrt{2}}
=  2   \, p^{[\mu } \varepsilon^{\nu ]}_{a  \dot a}   \,,
\ee
where $\Gmat^{\mu\nu} =  \frac{1}{2} (\Gmat^{\mu }\Gmat^{\nu }- \Gmat^{\nu }\Gmat^{\mu })$ are the rank-2 elements of the Clifford algebra as usual, and the antisymmetrization includes a factor of $1/2$.

In the massless limit,
the four on-shell vector states in Eq.~\eqref{massivepol} split into three symmetric and one antisymmetric one with respect to the diagonal little group. 
The latter mode is divergent in the massless limit,
\begin{equation}
\frac{m}{2}(\varepsilon^\mu_{a  \dot b} (p)-\varepsilon^\mu_{b  \dot a} (p))\Big|_{m \rightarrow 0 ,   \dot a \rightarrow a,  \dot b \rightarrow b} = ~ - \frac{1}{\sqrt{2}} p^\mu  \epsilon_{ab} \ .
\end{equation}
The symmetric states reduce to the massless polarization vector in Eq. (\ref{masslesspol}),
\begin{equation}
  \frac{1}{2}\big(\varepsilon^\mu_{a  \dot b} (p)+\varepsilon^\mu_{b  \dot a} (p)\big) \Big|_{m \rightarrow 0 ,   \dot a \rightarrow a, \dot b \rightarrow b}= ~  \varepsilon^\mu_{a  b} (k,q) \ .
\end{equation} 
The linearized field strength for the massive vector has no mass factor in the denominator, so it has a smooth massless limit.

In giving explicit representations for amplitudes in 5D, it will be convenient to dress the little-group indices with (Grassmann-even) auxiliary  variables $\auxu^a$, $\auxv^{\dot a}$, so that external states carry dressed polarization vectors in the massless and massive cases, 
\begin{equation}
\varepsilon^\mu_i = \varepsilon^\mu_{ab}(k_i,q_i)  \auxu^{a}_i \auxu^{b}_i \ , \qquad \qquad
\bep^\mu_i = \varepsilon^\mu_{a \dot b}(p_i)  \auxu^{a}_i \auxv^{\dot b}_i \ ,
\end{equation}
where we also introduced particle labels $i$.
An added benefit of this notation is that the auxiliary variables take care of the symmetrization of the little-group indices for the massless polarization.

\subsection{Tensor polarizations}

Five dimensions is the lowest dimension in which theories can exhibit asymptotic states corresponding to massive (anti)self-dual tensor fields~\cite{Townsend:1983xs},
\be
B_{\mu\nu} = \pm
\frac{i}{3! m} \epsilon_{\mu\nu\rho\sigma\lambda} 
H^{\rho\sigma\lambda} \ ,
\label{SD_ADS_position_space}
\ee
where $H_{\mu\nu\rho} = \partial_{[\mu}B_{\nu\rho]}$ is the field strength of the anti-symmetric tensor $B_{\mu\nu}$ and the 5D Levi-Civita symbol is normalized as $\epsilon_{01234}=\epsilon^{01234}=1$. For an on-shell tensor with $p^2 = m^2$ and $\partial^\mu B_{\mu\nu}=0$, the massive (anti)self-duality relation can also be written as
\be
\partial^\rho H_{\rho \mu \nu}=\mp  \frac{im}{3!}\epsilon_{\mu\nu \rho \sigma \lambda} H^{\rho \sigma \lambda}   \ .
\ee
The variables introduced in previous sections allow us to construct polarization tensors obeying all the required physical constraints. 
For a massive self-dual and anti-self-dual tensors they are
\begin{align}
\varepsilon^{\mu\nu}_{a  b} (p) =  \frac{ \langle {\bm p}_{a}| \Gmat^{\mu \nu }  | {\bm p}_{b} \rangle}{4\sqrt{2}m}\ , 
\qquad \qquad
\varepsilon^{\mu\nu}_{\dot a  \dot b} (p) =  \frac{ [{\bm p}_{\dot a}| \Gmat^{\mu \nu }  | {\bm p}_{\dot b} ]}{4\sqrt{2}m}\, ,
\label{poltensors}
\end{align}
respectively. These polarization tensors are transverse,
\be
p_\mu  \varepsilon^{\mu\nu}_{ab} (p)=0
~~,\qquad \qquad
p_\mu  \varepsilon^{\mu\nu}_{\dot a\dot b} (p)=0 \ .
\ee
They also satisfy the little-group completeness relation for each $SU(2)$ factor of the little group,
\bea
\varepsilon^{\mu\nu}_{\dot a  \dot b} (p)   \varepsilon_{\mu \nu,\dot c  \dot d} (p) &=& \frac{1}{2}(\epsilon_{\dot a \dot c}  \epsilon_{\dot b \dot d}+\epsilon_{\dot a \dot d}  \epsilon_{\dot b \dot c})\,, \nn \\
\varepsilon^{\mu\nu}_{ a b} (p)   \varepsilon_{\mu \nu,c d} (p) &=& \frac{1}{2}(\epsilon_{ac}  \epsilon_{bd}+\epsilon_{ad}  \epsilon_{bc})\, ,
\eea
as well as the Minkowski-space completeness relation in the space of transverse two-tensors,
\bea
\varepsilon^{\mu\nu}_{a  b} (p) \varepsilon^{\rho \sigma,a b} (p) &=& \frac{1}{4}(\tilde\eta^{\mu \rho } \tilde\eta^{\sigma \nu }- \tilde\eta^{\mu \sigma} \tilde\eta^{\rho \nu }) - \frac{\epsilon^{\mu\nu \rho \sigma \lambda} p_\lambda }{4m}\equiv {\cal P}_{\rm SD}\,, \nn \\
\varepsilon^{\mu\nu}_{\dot a \dot  b} (p) \varepsilon^{\rho \sigma, \dot a \dot b} (p) &=& \frac{1}{4}(\tilde\eta^{\mu \rho  } \tilde\eta^{\sigma \nu }- \tilde\eta^{\mu \sigma } \tilde\eta^{\rho  \nu }) + \frac{\epsilon^{\mu\nu \rho \sigma \lambda} p_\lambda }{4m}\equiv {\cal P}_{\rm ASD}\,.
\eea
Here ${\tilde\eta}^{\mu\nu}$
is the massive-vector physical-state projector defined in \eqn{massivephysproj}, 
and ${\cal P}_{\rm SD}$ and ${\cal P}_{\rm ASD}$ are the (anti-)self-dual tensor projectors. The latter satisfy $({\cal P}_{\rm SD})^2={\cal P}_{\rm SD}$, $({\cal P}_{\rm ASD})^2={\cal P}_{\rm ASD}$ and ${\cal P}_{\rm SD}{\cal P}_{\rm ASD}=0$.

The polarization tensors \eqref{poltensors} satisfy the momentum-space form of 
the (anti)self-duality relations in Eq.~\eqref{SD_ADS_position_space},
\bea
p^\rho    \epsilon_{\mu\nu \rho \sigma \lambda} \,\varepsilon^{\mu\nu}_{a  b}(p) &=& - 2m\, \varepsilon_{\sigma \lambda,a  b}(p)\,, \nn \\
p^\rho    \epsilon_{\mu\nu \rho \sigma \lambda} \,\varepsilon^{\mu\nu}_{\dot a  \dot b}(p) &=& 2m\, \varepsilon_{\sigma \lambda, \dot a  \dot b}(p)\,.
\eea
The polarization tensors in Eq. \eqref{poltensors} can also be obtained from the massive polarization vectors we have constructed through the relations
\bea
\varepsilon^{\mu\nu}_{a  b} (p) &=& - \frac{1}{\sqrt{2}}\varepsilon^{[\mu }_{a  \dot a} (p)   \varepsilon^{\nu] }_{b  \dot b} (p) \epsilon^{ \dot a  \dot b} \ ,  \nn \\
\varepsilon^{\mu\nu}_{\dot a  \dot b} (p) &=& \frac{1}{\sqrt{2}}\varepsilon^{[\mu }_{a  \dot a} (p)   \varepsilon^{\nu] }_{b  \dot b} (p) \epsilon^{a  b}  \ .
\eea
This  provides one way of realizing massive tensors in a gravitational theory using the double copy of massive vectors from a gauge theory. 

The massless limit of the polarization tensors \eqref{poltensors} is singular because of the manifest factors of $m^{-1}$; one can however dualize them to  polarization vectors which are finite  in the massless limit. For example, the self-dual tensor becomes
\be
\varepsilon^{\mu'}_{a  b}(k,q) =  2\varepsilon^{\mu\nu}_{a  b}(p) k^\rho   q^\sigma  \epsilon_{\mu\nu \rho \sigma \lambda} \eta^{\lambda  \mu'} \Big|_{m \rightarrow 0}\,,
\ee
where $p^\mu=k^\mu+m^2q^\mu$ as before, and the polarization vector is identical to the one introduced in Eq.~\eqref{masslesspol}.  For the anti-self-dual tensor $\varepsilon^{\mu\nu}_{\dot a  \dot  b}(p)$, the same limit holds up to an overall minus sign. In fact, the self-dual and anti-self-dual tensor can be combined into a tensor that is finite in the massless limit,
\be
\varepsilon^{\mu\nu}_{ab} (k,q)= \varepsilon^{\mu\nu}_{a  b}(p)- \varepsilon^{\mu\nu}_{\dot a  \dot b}(p)\Big|_{\dot a, \dot b \rightarrow a,b;\, m \rightarrow 0} \ .
\ee
This corresponds to the physical state of a massless vector, albeit written as tensor. 

Combining the above relations we have the following (non-linear) relation for the massless polarization  vector,
\be
\varepsilon^{\mu'}_{a  b}(k,q) = -  \sqrt{2} \varepsilon^{\mu}_{a  c} (k,q)   \varepsilon^{\nu }_{b  d} (k,q) \epsilon^{c d}  k^\rho   q^\sigma  \epsilon_{\mu\nu \rho \sigma \lambda} \eta^{\lambda  \mu'} \ ,
\ee
which is equivalent to the statement that, in 5D, the double copy of two massless vectors contains a massless vector in its antisymmetric part. 

\subsection{Conformal generators}
\def\NcN{\cN}
In this subsection, we briefly comment on conformal symmetry in five and six dimensions, and their possible supersymmetric extensions. It is convenient to start in 6D, and infer details of the 5D case via dimensional reduction.   

The unitary representations of the 6D conformal group $SO(6,2)=SO^*(8)$  and their extensions to superconformal algebras $OSp(8^*|\NcN)$, with even $\NcN$, using twistorial oscillators were studied in Refs. \cite{Gunaydin:1984wc,Gunaydin:1999ci,Chiodaroli:2011pp}. 
The generators of $SO(6,2)$ were realized as bilinears of twistorial oscillators and one finds that the group admits infinitely many representations, referred to as doubletons, that describe massless conformal fields of ever increasing spins. These massless conformal fields correspond to symmetric tensors of the spinor representation of the 6D Lorentz group $SU^*(4)$.  For the doubletons, the Poincar\'e mass operator vanishes identically. 
Similarly, the conformal superalgebras $OSp(8^*|\NcN)$, with even subalgebras $SO^*(8) \oplus USp(\NcN)$, exist for any even $\cN$ and admit infinitely-many conformally-massless unitary supermultiplets of ever increasing spins. In six dimensions, coordinates and momenta can be described by anti-symmetric tensors of twistorial variables in coordinate-space or momentum-space pictures. The twistorial oscillators formulated in Ref.~\cite{Chiodaroli:2011pp} satisfy the commutation relations
\def\etaK{\xi}
\begin{equation}
    [\etaK_{\hat{A}a},\lambda^{\hat{B}b}]= \frac{1}{2} \delta^{\hat{B}}_{\hat{A}} \delta^b_a  \ , 
\end{equation}
where $\hat{A} , \hat{B}=1,2,3,4$ are the spinor representation indices of $SU^*(4)$ and $a,b=1,2$. 
Then, the translation $P_{\hat{\mu}}$ and  special-conformal generators $K_{\hat{\mu}}$ in six dimensions can be represented as 
\begin{equation}
P^{\hat{A}\hat{B}}= - \frac{1}{2} (\Sigma^{\hat{\mu}})^{\hat{A}\hat{B}} P_{\hat{\mu}} = \lambda^{\hat{A}a} \lambda^{\hat{B}b} \epsilon_{ba} \ ,\quad  \quad 
K_{\hat{A}\hat{B}}=-\frac{1}{2}(\Sigma^{\hat{\mu}})_{\hat{A}\hat{B}} K_{\hat{\mu}} =  \etaK_{\hat{A}a} \etaK_{\hat{B}b} \epsilon^{ba} \ ,
\end{equation}
where  $\Sigma^{\hat{\mu}}$ are the anti-symmetric 6D sigma matrices and  $\hat{\mu},\hat{\nu},\ldots=0,1,\ldots,5$. Under commutation, they close into the Lorentz group $SU^*(4)$ generators $M^{\hat{A}}_{\ \ \hat{B}}$ and dilatation generator $\mathcal{D}$.

Under the 5D Lorentz group $USp(2,2)$, the 6D translation and special conformal generator  decompose as $5+1$ which corresponds to taking the symplectic traces in the antisymmetric tensor representation of $SU^*(4)$,
\begin{eqnarray}
    P^{\hat{A}\hat{B}} = P^{AB} + \frac{1}{4} \Omega^{AB} P \ ,  &\qquad &
    K_{\hat{A}\hat{B}} = K_{AB} + \frac{1}{4} \Omega_{AB} K  \ ,
\end{eqnarray}
where as before $\Omega_{AB}$ is the $USp(2,2)$ invariant symplectic metric, and 
\begin{eqnarray}
    P^{AB}= P^{\hat{A}\hat{B}} -\frac{1}{4} \Omega^{AB} \,  P \ , & \qquad&      P= \frac{1}{4} \Omega_{AB} P^{\hat{A}\hat{B}} = P_5   \ ,  \\
    K_{AB} = K_{\hat{A}\hat{B}} - \frac{1}{4} \Omega_{AB} \, K \ , &&
    K = \frac{1}{4} \Omega^{AB}  K_{\hat{A}\hat{B}} \ .
\end{eqnarray}
The generators $P^{AB}$ and $K_{AB}$ close into the $USp(2,2)$ generators $M^A_{\ \  B}$ and dilatation generator. Together, they generate the 5D conformal group $SO(5,2)$. The conformally-massless representations of $SO(6,2)$ lead to massive representations of $SO(5,2)$ under the above reduction since
\begin{equation}
    P^{\hat{\mu}} P_{\hat{\mu}} =0~~~~ \Longrightarrow ~~~~ P^\mu P_\mu = (P_5)^2 = M^2 \ ,
\end{equation}
where $M$ is the mass. Restricting to the massless $SO(5,2)$ representations is not possible without modifying the oscillator construction. It follows from general results that conformal groups $SO(D,2)$ in even dimensions $D$ admit infinitely many unitary representations describing massless conformal fields of ever increasing spins, while, in odd dimensions, one finds only two unitary representations that describe massless conformal fields, namely scalar and spinor fields \cite{Angelopoulos:1997ij}. 
At a deeper level, this follows from the following fact. Massless conformal fields in any dimension \cite{Fernando:2015tiu} are described by the so-called minimal unitary representations of the conformal group $SO(D,2)$ and their deformations. A true minimal unitary representation of $SO(D,2)$ describes a massless conformal scalar in $D$ dimensions. This minimal unitary representation admits infinitely many deformations labeled by the quadratic Casimir of the little group $SO(D-2)$ of massless particles in even dimensions. They describe massless conformal fields of ever increasing spins. On the other hand, in odd dimensions, the minimal unitary representation of the conformal group  admits a single deformation that describes a massless spinor field~\cite{Fernando:2015tiu}. Therefore, in five dimensions, only the scalar and spinor fields can be conformally massless~\cite{Fernando:2014pya}. The so-called two remarkable representations of the 3D conformal group $Sp(4,\mathbb{R})= SO(3,2)$ that were first studied by Dirac and labeled as  singletons correspond simply to the minimal unitary representation and its spinorial  deformation. 

The minimal unitary representation of the 5D conformal group $SO(5,2)$ and its deformation as well as their supersymmetric extension were studied in Refs. \cite{Fernando:2014pya,Gunaydin:2016amv}.
There exists a unique simple superconformal algebra in five dimensions, namely the exceptional superalgebra $F(4)$ with the even subalgebra $SO(5,2)\oplus SU(2)$~\cite{Nahm:1977tg}. The minimal unitary supermultiplet of $F(4)$ consists of two complex scalar fields in the doublet of R-symmetry group $SU(2)_{\rm R}$  and a symplectic-Majorana spinor, which are both conformally massless \cite{Fernando:2014pya}. 
The superalgebra $F(4)$ is not a subalgebra of any of the 6D superconformal algebras $OSp(8^*|\NcN)$.

In this paper we will not study the $F(4)$ superalgebra, however we will study the massive representations that come from 6D superconformal algebra. In particular, the Poincar\'e subalgebra of 6D superconformal algebra $OSp(8^*|\NcN)$ descends directly to $\NcN$ extended Poincar\'e superalgebra in 5D with the momentum generator in the fifth spatial dimension acting as a singlet central charge under the R-symmetry group $USp(\NcN)$.

\section{Supersymmetry\label{sec3}}
 
We now consider the Poincar\'e superalgebras in five dimensions. $\cal N$-extended  Poincar\'e superalgebras with central charges have the general form~\cite{Hull:2000cf}
\be
\{Q^I_A, Q^J_B\} = {\bf \Omega}^{IJ}\,  (\Gmat^\mu )_{AB} P_\mu +  \Omega_{AB} \,(Z^{I J} + Z \, {\bf \Omega}^{IJ} ) \ ,
\ee
where the supercharge $Q^I_A$ carries a lower $USp(2,2)$ Lorentz index and an upper $USp({\cal N})$ R-symmetry index. The  $ {\bf \Omega}^{I J}$ is the symplectic metric of $USp({\cal N})$, and the central charge is here decomposed into a singlet $Z$ and non-singlet $Z^{IJ}$ of this group. The non-singlet central charge $Z^{IJ}$ is antisymmetric and ${\bf \Omega}$-traceless. Massless representations of the Poincar\'e superalgebras have vanishing central charges. 

In this section, we discuss in detail the realization of both the massless and massive cases for (minimal) $\cN=2$ and $\cN=4$ supersymmetry.\footnote{In our convention, the number of supercharges is $4\cN$. For massive 1/2-BPS multiplets we use a label $(r_+,r_-)$ to denote the number of chiral and anti-chiral little-group components of the supercharges, where $\cN=2(r_++r_-)$~\cite{Hull:2000cf}.
} 
This is sufficient for describing super-Yang-Mills amplitudes and theories that have tensor multiplets. For gravitational theories, which may have up to $\cN=8$ supersymmetry, we will infer the details via the double copy.  
All relevant 1/2-BPS on-shell superfields will be given, and their double-copy relations will be exhibited.

\subsection{$\cN=2$ supersymmetry without central charge}

\def\qsusy{\mathfrak{q}}
\def\qsusym{\bm{\mathfrak{q}}}

The  ${{\cal N} = 2}$ Poincar\'e superalgebra in five dimensions without central charges is given by
\be
\{Q^\alpha_A, Q^\beta_B\} = \epsilon^{\alpha \beta}\,  
\Gmat^\mu_{AB} P_\mu \,  ,
\label{susyalg}
\ee
where $\alpha,\beta$ are $SU(2)$ R-symmetry indices and $A$ and $B$ are as before the fundamental $USp(2,2)$ indices. The supercharge for a one-particle state of null momentum $p_\mu$ can be written as a product of an on-shell spinor, $| p^a \rangle_A$, and Grassmann-odd variables~$\theta_{a}^{\alpha}$, 
\be
Q^\alpha_A \Big|_\text{1-pt}= | p^a \rangle_A \theta_{a}^{\alpha} \equiv    | \qsusy^\alpha \rangle_A  \, ,
\ee
where we  defined the Grassmann-odd symplectic-Majorana spinor $ | \qsusy^\alpha \rangle_A$. In term of this spinor, it follows that $\theta^\alpha_a = \langle q_a | \mathfrak{q}^\alpha \rangle$, where $\langle q_a|$ is the usual reference spinor satisfying $\langle q_a | p_b \rangle =\epsilon_{ab}$.

As a consequence of \eqref{susyalg},  $\theta_{a}^{\alpha}$ must satisfy the supersymmetry algebra projected on the little group, 
\be
\{\theta^\alpha_a, \theta^\beta_b\} = 
-\epsilon^{\alpha \beta}  \epsilon_{ab} \ .
\label{littlegroupsusy}
\ee
The possible solutions of Eq.~\eqref{littlegroupsusy} are related  by $SU(2)\times SU(2)$ transformations,\footnote{ Naively, 
the right-hand side of this relation is invariant under two independent $SL(2,\mathbb{C})$ transformations, one acting on the indices $a,b, ...$  and one acting on the R-symmetry group  indices $ \alpha , \beta ,..$. However R-symmetry group is always compact and the indices $a,b,..$  refer to the little-group $SU(2)$ indices inside  $SL(2,\mathbb{C})$.   }
\be \label{oscillatorSol}
\theta'^\alpha_a =  E_{a}^{\ b} \, \theta^\beta_b\,\widetilde  E_\beta^{\ \alpha}\,,
\ee
where ${\rm Det}\,E={\rm Det}\,\tilde E =1 $. For our purpose, it is convenient to choose a solution that maintains little-group covariance, at the expense of manifest R-symmetry.
An explicit parametrization of $\theta_a^{\alpha}$ that satisfies this algebra is given by the oscillator representation 
\be
\theta^{ \alpha}_a= (\theta_a)^\alpha = \left(\begin{matrix}
\eta_a  \\
{-}\frac{\partial}{\partial \eta^a}  \\
\end{matrix}\right) \ ,
\ee
where $\eta_a$ are two unconstrained Grassmann-odd auxiliary parameters, which together transform as a spinor in the little group and a scalar in the Lorentz group. We may also assign a $U(1)$ charge of $1/2$ to $\eta_a$, and charge $-1/2$ to $\frac{\partial}{\partial \eta^a}$, which corresponds to the Cartan generator of the broken $SU(2)$. Thus, the $\eta_a$ are complex variables.

Having defined the one-particle supercharge, it follows that the multi-particle supercharge is the sum
\be
Q^\alpha_A  = \sum_{i=1}^n  | \mathfrak{q}^\alpha_i \rangle_A \,,
\ee
where $i$ is the particle label. It is easy to check the full supersymmetry algebra,
\be
\{Q^\alpha_A, Q^\beta_B\} = \sum_{i,j=1}^n \{  | \mathfrak{q}^\alpha_i \rangle_A ,  | \mathfrak{q}^\beta_j \rangle_B \} =  \sum_{i,j=1}^n  | p^a_i \rangle_A  | p^b_j \rangle_B   \{ \theta^\alpha_{ai} ,  \theta^\beta_{b j} \} =\epsilon^{\alpha \beta}\,  \Gmat^\mu_{AB} P_\mu \, ,
\ee
where we used that $\{ \theta^\alpha_{ai} ,  \theta^\beta_{a j} \} = {-}\delta_{ij}\,\epsilon^{\alpha \beta}  \epsilon_{ab}$, since the little group of each particle is independent, and the total momentum is defined as $P^\mu= \sum_{i} p_i^\mu $.

\subsection{$\cN = 2$ supersymmetry  with central charge} \label{Neq2susySection}

The  ${{\cal N} = 2}$ supersymmetry algebra in 5D admits a singlet 
central charge,
\be
\{Q^\alpha_A, Q^\beta_B\} = \epsilon^{\alpha \beta}\,  \Gmat^\mu_{AB} P_\mu +Z \epsilon^{\alpha \beta}\,  \Omega_{AB} \,.
\ee
On dimensional grounds, it is possible to identify the central charge with mass, hence we rewrite it as
\be
Z= M \,.
\ee
We can repeat the construction of the single-particle supercharge as in the massless case above, while replacing the massless spinor with a massive one. For a one-particle state of momentum $p_\mu$ and mass $m$, we now use a massive spinor to extract the little-group dependence,
\be
Q^\alpha_A \Big|_\text{1-pt} =   | {\bm p}^a \rangle_A \theta_{a}^{\alpha} \equiv   |\qsusym^\alpha  \rangle_A \, .
\ee
This implies that the supercharges can now be  chiral in the sense that they only involve one of the $SU(2)$ factors of the little group $SO(4) \sim SU(2)\times SU(2)$.
The fermionic oscillator algebra is now chiral in the same sense,
\be
\{\theta^\alpha_a, \theta^\beta_b\} = {-}\epsilon^{\alpha \beta}  \epsilon_{ab}\, .
\ee
 As before, this relation has a family of solutions parameterized by $SU(2)\times SU(2)$; a representative, written in terms of an unconstrained Grassmann variable and the corresponding derivative, is
\be
\theta^{ \alpha}_a= (\theta_a)^\alpha = \left(\begin{matrix}
\eta_a  \\
{-}\frac{\partial}{\partial \eta^a}  \\
\end{matrix}\right) \, ,
\ee
where, as in the massless case, we have chosen to break the $SU(2)$ R-symmetry down to a manifest $U(1)$, while maintaining the $SU(2)$ little-group symmetry.
The multi-particle supersymmetry algebra, whose generators are sums of single-particle supersymmetry generators, follows straightforwardly,
\be
\{Q^\alpha_A, Q^\beta_B\} = \sum_{i,j=1}^n \{  | \qsusym^\alpha_i \rangle_A ,  | \qsusym^\beta_j \rangle_B \} =  \sum_{i,j=1}^n  | {\bm p}^a_i \rangle_A  | {\bm p}^b_j \rangle_B   \{ \theta^\alpha_{ai} ,  \theta^\beta_{b j} \} =\epsilon^{\alpha \beta}(  \Gmat^\mu_{AB} P_\mu + \Omega_{AB} M) \, .
\ee
The last equality follows from \eqn{spinorCompl}, $P^\mu= \sum_{i} p_i^\mu$ and $M= \sum_{i} m_i$. 

\subsection{${{\cal N}=4}$ supersymmetry without central charge}

The massless ${{\cal N} = 4}$ supersymmetry algebra in 5D is
\be
\{Q^{\dot A}_A, Q^{\dot B}_B\} = \Omega^{\dot A \dot B}\,  \Gmat^\mu_{AB} P_\mu \,,   
\ee
where $\dot A, \dot B$ are $USp(4)$ R-symmetry indices.\footnote{Since the R-symmetry group $USp(4) \cong SO(5)$ only differs from the Lorentz group $USp(2,2) \cong SO(1,4)$ by a signature change, it is convenient recycle the notation by putting dots on all indices. Specifically, we take $\Omega_{\dot A \dot B}$ to be given by \eqn{OmegaDef}.} As before, the supercharge for a one-particle state of null momentum $p_\mu$ can be written by factoring out an on-shell spinor carrying the Lorentz index,
\be
Q^{\dot A}_A \Big|_\text{1-pt}= | p^a \rangle_A \theta_{a}^{\dot A} \equiv   | \qsusy^{\dot A} \rangle_A  \,,
\ee
where $\theta_{a}^{\alpha}$ are Grassmann-odd and satisfy the supersymmetry algebra projected on the little group, 
\be \label{Neq4littlegroupSUSY}
\{\theta^{\dot A}_a, \theta^{\dot B}_b\} = {-}\Omega^{\dot A \dot B}  \epsilon_{ab}\,.
\ee
An explicit parametrization of $\theta_a^{\alpha}$ that satisfies the algebra is given by the oscillator representation 
\be \label{BadNeq4oscillatorRep}
\theta^{\dot A}_a= (\theta_a)^{\dot A} = \left(\begin{matrix}
\theta_a^{1 \alpha }  \\
\theta_a^{2\alpha}   \\
\end{matrix}\right)\, ~~~{\rm with}~~~~
\theta_a^{\hat \imath \alpha}= \left(\begin{matrix}
\hat \eta_a^{\hat \imath}  \\
{-}\frac{\partial}{\partial \hat \eta^a_{\hat \imath} }  \\
\end{matrix}\right)\,,
\ee
where we used the $SO(2)\times SU(2)$ subgroup of $SO(5)$ by writing the four-component $USp(4)$ index as a product of two fundamental indices, $\dot A= \hat \imath \otimes \alpha$. Alternatively, we can write a solution that does not break the diagonal $SU(2)$ factor,\footnote{Note that a simpler $SU(2)$-preserving parametrization of $\theta^{\dot A}_a$ can be obtained after doing a similarity transform on $\Omega^{\dot A \dot B}$, but here we work with the block-diagonal symplectic metric~(\ref{OmegaDef}).} 
\be \label{Neq4oscillatorRep}
\theta^{\dot A}_a= (\theta_a)^{\dot A} = \frac{i}{\sqrt{2}} \left(\begin{matrix}
\eta^\alpha_a + \frac{\partial}{\partial \eta^a_\alpha}   \\
\eta^\alpha_a - \frac{\partial}{\partial \eta^a_\alpha}  \\
\end{matrix}\right)\, .
\ee
This solution gives the maximal R symmetry that can be realized in amplitudes without breaking little-group symmetry. The unconstrained Grassmann-odd auxiliary parameters $\eta_a^\alpha$ have four complex components, making the $SU(2)$ little group and $ U(1) \times SU(2) \subset USp(4)$ R-symmetry subgroup manifest.  

Similarly to previous cases, the multi-particle supersymmetry algebra follows by summing over particle labels $i,j$,
\be
\{Q^{\dot A}_A, Q^{\dot B}_B\} = \sum_{i,j=1}^n \{  | \qsusym^{\dot A}_i \rangle_A ,  | \qsusym^{\dot B}_j \rangle_B \} =  \sum_{i,j=1}^n  | {\bm p}^a_i \rangle_A  | {\bm p}^b_j \rangle_B   \{ \theta^{\dot A}_{ai} ,  \theta^{\dot B}_{b j} \} =\Omega^{\dot A \dot B}  \Gmat^\mu_{AB} P_\mu  \,.
\ee

\subsection{$\mathcal{N}=4$ supersymmetry with central charge \label{sec:Neq4WCentralCharge}}

The massive ${{\cal N} = 4}$ supersymmetry algebra in 5D includes an antisymmetric central charge $Z^{\dot A \dot B}$,
\be
\{Q^{\dot A}_A, Q^{\dot B}_B\} = \Omega^{\dot A \dot B}\,  \Gmat^\mu_{AB} P_\mu +Z^{\dot A \dot B}\,  \Omega_{AB} \,. 
\ee
We will first consider the case in which the central charge is a singlet of R-symmetry group $USp(4)$.  Thus, it is of the form
\be
Z^{\dot A \dot B} = M\, \Omega^{\dot A \dot B}\,,
\ee
which is a direct extension of the ${\cal N}=2$ case in Section~\ref{Neq2susySection}. We will construct a supercharge which only depend on the massive angle spinor,
\be
Q^{\dot A}_A \Big|_\text{1-pt} =  | {\bm p}^a \rangle_A \theta_{a}^{\dot A}   \equiv    | \qsusym^{\dot A}  \rangle_A\,.
\ee
Thus, in this case, the supersymmetry is chiral, which, in 6D language, corresponds to $(2,0)$ supersymmetry.

The Grassmann-odd parameters satisfy the same little-group superalgebra as in the massless case (\ref{Neq4littlegroupSUSY}),
\be
\{\theta^{\dot A}_a, \theta^{\dot B}_b\} = -\Omega^{\dot A \dot B}  \epsilon_{ab}\,.
\ee
The explicit solution for this algebra is then identical to \eqn{Neq4oscillatorRep}. This is not surprising, since the little-group and R symmetries are identical to the massless case. The full algebra is then
\be
\{Q^{\dot A}_A, Q^{\dot B}_B\} = \sum_{i,j=1}^n \{  | \qsusym^{\dot A}_i \rangle_A ,  | \qsusym^{\dot B}_j \rangle_B \} =  \sum_{i,j=1}^n  | {\bm p}^a_i \rangle_A  | {\bm p}^b_j \rangle_B   \{ \theta^{\dot A}_{ai} ,  \theta^{\dot B}_{b j} \} =\Omega^{\dot A \dot B}  (\Gmat^\mu_{AB} P_\mu + M \Omega_{AB})  \,,
\ee
where the mass term $M= \sum_{i} m_i$ comes from the completeness relation in \eqn{spinorCompl}.

Next, consider the case in which $Z^{\dot A \dot B}$ is not a singlet and, thus, breaks $USp(4)$. We start with the simplifying assumption that the central charge is proportional to a $SO(5)$ gamma matrix. For example, taking $i\Gamma^9$ gives the block-diagonal form
\be
Z^{\dot A \dot B} = i M\, (\Gamma^9)^{\dot A \dot B} =
M \left(\begin{matrix}
\epsilon^{\alpha \beta} & 0 \\
0&  -\epsilon^{\dot \alpha \dot \beta}   \\
\end{matrix}\right)\,.
\ee
Here we take the $SO(5)$ index to run from $5$ to $9$, and the internal $SO(5)$ gamma matrices are simply related to the spacetime ones by a Wick rotation,
\be
\Gamma^{n}_{\dot A \dot B}= \Big({-}i\gamma^{0}_{AB},~ \gamma^{\mu>0}_{AB} \Big)\Big|_{A\rightarrow \dot A,B\rightarrow \dot B,n=\mu+5}\,.
\ee

With the central charge pointing in a fixed direction, the R-symmetry is broken to $SO(4) \sim SU(2)\times SU(2)$, and the supercharges separate into those that transform in each copy of $SU(2)$, 
\be
Q^{\dot A}_A \rightarrow
\left(\begin{matrix}
Q^{\alpha}_A  \\
\tilde Q^{\dot \alpha}_A  \\
\end{matrix}\right) \,,
\ee
where, as before, the Greek indices are R-symmetry and the dot distinguishes the left and right $SU(2)$ factors.
For a one-particle state of momentum $p_\mu$ and mass $m$, we use the massive spinors to extract the little-group dependence,
\bea
Q^{\alpha}_A \Big|_\text{1-pt}&=& | {\bm p}^a \rangle_A \theta_{a}^{\alpha}   \equiv    | \qsusym^{\alpha}  \rangle_A  \,, \nn \\
\tilde Q^{\dot \alpha}_A \Big|_\text{1-pt}&=&  | {\bm p}^{\dot a} ]_{A} \tilde \theta^{\dot \alpha}_{\dot a}   \equiv    | \qsusym^{\dot \alpha}  ]_A  \, .
\eea
We are led to two independent oscillator algebras
\be \label{twoOscillators}
\{\theta^{\alpha}_a, \theta^{\beta}_b\} = {-}\epsilon^{\alpha \beta}  \epsilon_{ab}\,,~~~~~~~ 
\{\tilde \theta^{\dot \alpha}_{\dot a}, \tilde\theta^{\dot \beta}_{\dot b}\} = {-}\epsilon^{\dot \alpha \dot \beta}  \epsilon_{\dot a \dot b}\,,~~~~~~~  
\{\theta^\alpha_a, \tilde \theta^{\dot \beta}_{\dot b}\}=0\,,
\ee
which are solved by
\be \label{twoOscillatorsSol}
\theta^{\alpha}_a= (\theta_a)^{\alpha} = \left(\begin{matrix}
\eta_a  \\
{-}\frac{\partial}{\partial \eta^a}  \\
\end{matrix}\right) \,,
~~~~~~~~
\tilde \theta^{\dot \alpha}_{\dot a}= (\tilde\theta_{\dot a})^{\dot \alpha} = \left(\begin{matrix}
 \etaT_{\dot a}  \\
{-}\frac{\partial}{\partial  \etaT^{\dot a}}  \\
\end{matrix}\right) \,,
\ee
where now we have two sets of complex unconstrained auxiliary Grassmann parameters $\eta^{a}$, $ \etaT^{\dot a}$. They make the little group $SU(2)\times SU(2)$ manifest and break the R-symmetry down to $U(1)\times U(1) \subset SU(2)\times SU(2)$. Note that this is isomorphic to the solution we wrote down in~\eqn{BadNeq4oscillatorRep}.

Now consider the more general case where the central charge is a generic $SO(5)$ vector,\footnote{Antisymmetry rules out higher-rank elements of the Clifford algebra in the decomposition of the central charge. Also, for 1/2-BPS multiplets in super-Yang-Mills theory, the singlet and $SO(5)$-vector central charge are mutually exclusive~\cite{Hull:2000cf}.} 
\be
Z_{\dot A \dot B} =  i \Gamma^n_{\dot A \dot B} M_n \ .
\ee
On general grounds, we expect that we need 10D Majorana-Weyl spinors to construct single-particle supercharges, and that we should think of a ten-dimensional momentum $p^{[10]}$ as decomposed as $p^{[10]}_{N}=(p_\nu, \mu_n)$, where $p_\nu=(p_0,p_1,p_2,p_3,p_4)$ is $SO(1,4)$ momentum, and $\mu_n=(\mu_5,\mu_6,\mu_7,\mu_8,\mu_9)$ is $SO(5)$ momentum. For $p^{[10]}$ to be massless, the 5-vectors must satisfy the constraint $p^2-m^2=0$ and $\mu^2+m^2=0$ (with negative $SO(5)$ signature such that $\mu^2<0$). For a generic multi-particle state the  mass vector appearing in the central charge is given as $M^n = \sum_j \mu_j^n$, where $j$ labels the particles.

We can then obtain new spinors by tensoring the 5D massive spinors as
\be
\lambda^{a \alpha}_{A \dot A} \equiv \frac{1}{\sqrt{2m}} | {\bm p}^a \rangle_A  | {\bm \mu}^{ \alpha}\rangle_{\dot A} \qquad 
{\rm and} 
\qquad \tilde \lambda^{\dot a \dot \alpha}_{A \dot A} \equiv \frac{1}{\sqrt{2m}} | {\bm p}^{\dot a} ]_A  | {\bm \mu}^{\dot \alpha} ]_{\dot A} \ ,
\label{twospinorproducts}
\ee
where the $SO(5)$ spinors can be obtained from \eqn{USp4spinorMassive} by an appropriate Wick rotation:\footnote{The insertion of $i$'s are chosen such that the completeness relations for the spinors are 
$
| {\bm \mu}_a\rangle_{\dot A}  \langle {\bm \mu}^a |^{\dot B} =
i\mu_{{\dot A}}^{\ {\dot B}} + m\, \delta_{{\dot A}}^{\dot B}
$ 
and 
$
| {\bm \mu}_{\dot a}]_{\dot A}  [ {\bm \mu}^{\dot a}|^{\dot B} = 
-i\mu_{{\dot A}}^{\ {\dot B}} + m\, \delta_{{\dot A}}^{\dot B} 
$.
} $| {\bm \mu} \rangle = | {\bm p} \rangle\Big|_{p^{j>0} \rightarrow i \mu^{j+5}} $ and $| {\bm \mu}] = i | {\bm p}]\Big|_{p^{j>0} \rightarrow i \mu^{j+5}} $. The resulting 10D Majorana-Weyl spinor is the direct sum of the above spinors, and it has $4+4=8$ degrees of freedom, as is manifest from inspecting the little-group indices.   

The one-particle supercharge can now be written more covariantly as 
\be
Q_{A}^{\dot A} \Big|_\text{1-pt} = \frac{1}{\sqrt{2m}} \Big( | {\bm p}^a \rangle_A  \theta^{\alpha}_a  \langle {\bm \mu}_{\alpha} |^{\dot A} +| {\bm p}^{\dot a} ]_A  \tilde \theta^{\dot \alpha}_{\dot a}  [ {\bm \mu}_{ \dot \alpha}|^{\dot A}  \Big)  \equiv |  \qsusym^{\dot A}\rangle_A \,,
\ee
where we may now think of $ |  \qsusym^{\dot A}\rangle_A  $ as either a 5D symplectic-Majorana spinor or a 10D Majorana-Weyl spinor.
The supersymmetry algebra now becomes
\bea \label{10Dsusy}
\{ Q_{A}^{\dot A},  Q_{B}^{\dot B} \} &=& \sum_{i,j=1} \{  | \qsusym^{\dot A}_i \rangle_A ,  | \qsusym^{\dot B}_j \rangle_B \}  \nn \\
&& \hskip -0.7in =  \sum_{i,j=1} \frac{1}{2m_i} \Big( | {\bm p}^a_i \rangle_A  | {\bm p}^b_j \rangle_B   \{ \theta^{\alpha}_{ai} ,  \theta^{ \beta}_{b j} \}   \langle {\bm \mu}_{\alpha i} |^{\dot A}  \langle {\bm \mu}_{\beta j} |^{\dot B}  + | {\bm p}^{\dot a}_i ]_A  | {\bm p}^{\dot b}_j ]_B   \{\tilde \theta^{\dot\alpha}_{\dot ai} ,  \tilde \theta^{\dot \beta}_{\dot b j} \}   [{\bm \mu}_{\dot \alpha i} |^{\dot A}  [{\bm \mu}_{ \dot \beta j} |^{\dot B} \Big)
\nn \\
&& \hskip-1.8cm =~
\Omega^{\dot A \dot B}  \Gmat^\mu_{AB} P_\mu + i \Omega_{AB} (\Gamma^n)^{\dot A \dot B} M_n  \, ,
\eea
where $\theta^{\alpha}_{ai}$ and $\tilde \theta^{\dot \alpha}_{\dot a i}$ obey the little-group supersymmetry algebras (\ref{twoOscillators}) with the same solution (\ref{twoOscillatorsSol}). Indeed, the two last cases we considered, namely the central charge proportional to $i \Gamma^9$ or to a general $SO(5)$ vector, are related to each other by a $SO(5)$ rotation, so it is not surprising that the little-group details are the same.

Let us briefly explain the appearance of the imaginary unit in front of the mass in the central charge. If follows from constructing the 10D gamma matrices, they are given by the following 32-by-32 matrices,
\be
(\mathbf{\Gamma}^{N})_{\cal A}^{\ \ {\cal B}} =
\left(\begin{matrix}
0 & (\mathbf{\Sigma}^{N})_{A \dot A}^{\ \ B \dot B}\\ 
 ({\overline{\mathbf{\Sigma}}}^{N})_{A \dot A}^{\ \ B \dot B} & 0 \\
\end{matrix}\right) \,,
\ee
where the 16-by-16 sigma matrices are $(\mathbf{\Sigma}^{N})_{A \dot A}^{\ \ B \dot B}= \Big( (\gamma^\nu)_{A}^{\ B} \delta_{\dot A}^{\ \dot B}~,~ i \delta_{A}^{\ B} (\Gamma^n)_{\dot A}^{\ \dot B} \Big)$ and $({\overline{\mathbf{\Sigma}}}^{N})_{A \dot A}^{\ \ B \dot B}= \Big( (\gamma^\nu)_{A}^{\ B} \delta_{\dot A}^{\ \dot B}~,~ -i \delta_{A}^{\ B} (\Gamma^n)_{\dot A}^{\ \dot B}\Big)$. From this we see that $i \Gamma^n$ is the natural expression appearing in the sigma matrices, and it is $\mathbf{\Sigma}^{N}$ that appears on the final line of the anti-commutator \eqn{10Dsusy}. Hence, the supersymmetry algebra  can be written more compactly as,
\be
\{ Q_{\dot A}^B,   Q^{\dot  B}_{A} \}=(\mathbf{\Sigma}^{N})_{A \dot A}^{\ \ B \dot B} P_N^{[10]}\,,
\ee
where $P_N^{[10]}=(P_\nu, M_n)$ is the 10D momentum generator.

\subsection{On-shell superfields for  massless 5D multiplets}

The unconstrained Grassmann variables $\eta$ are the building blocks of the 5D on-shell multiplets. We list here the on-shell superfields describing massless 1/2-BPS multiplets; the massive ones will be described in the next subsection. All displayed indices are either $SU(2)$ little-group indices (Latin), or $SU(2)$ R-symmetry indices (Greek). 

The on-shell vector supermultiplet with ${\cal N}=2$ supersymmetry is 
\be
{\cal V}^{{\cal N}=2}_a = \chi^1_a+ \Big(A_{a b} + {\epsilon_{ab} \over \sqrt{2}} \phi \, \Big) \eta^b+   \chi^2_{a} \, (\eta)^2  \ , 
\label{Neq2Vector}
\ee
where we have defined the square of the Grassmann variables as  $(\eta)^2={1 \over 2 }\eta_a \eta^a = \eta^1 \eta^2$. This superfield contains four bosonic\footnote{Imposing proper normalization for the scalar gives the $\sqrt{2}$ factor, since $\epsilon^{ab}/\sqrt{2}$ has unit norm. 
} ($A^{ab}{=}A^{ba},~ \phi $) and four fermionic $\chi_a^\alpha$ degrees of freedom, the latter are symplectic-Majorana spinors. 

The superfield transforms as a covariant $SU(2)$ little-group spinor, while the $SU(2)$ R-symmetry of the gaugino $\chi^\alpha_a$ is not manifest.
This is a consequence of the chosen oscillator representation of the supersymmetry algebra, which breaks manifest R-symmetry, and thus realizes linearly only half of the supercharges.\footnote{The choice here is analogous in spirit with the non-chiral 4D superspace of Ref.~\cite{Huang:2011um}.} Only a $U(1)$ remnant of the R-symmetry remains manifest in the charge carried by the $\eta$ and the fermions.

The hypermultiplet is a little-group singlet and an R-symmetry doublet,
\be
\Phi_{{\cal N}=2}^\alpha = {\phi\vphantom{\overline{\phi}}}^\alpha + \chi^\alpha_a \eta^a+  {\overline{\phi}}^\alpha (\eta)^2 \ ,
\ee
and contains four bosonic and four fermionic degrees of freedom. The vector and hypermultiplets are related through the double copy involving a fermion $\chi^\alpha_a$,
\be
{\cal V}_{{\cal N}=2}^a =\Phi_{{\cal N}=2}^\alpha \otimes \chi_\alpha^a\, ,
\ee
where the index contraction is needed to obtain the same number of degrees of freedom as in Eq.~\eqref{Neq2Vector}.

The graviton multiplet with ${\cal N}=2$ supersymmetry transforms as a spin-3/2 representation of the little group,
\be
{\cal H}^{{\cal N}=2}_{abc} = \psi^1_{abc} + \Big(h_{abcd} + {1  \over \sqrt{2}}A_{(a b} {\epsilon_{c)d} }  \Big) \eta^{d} + \psi^2_{abc} (\eta)^2 \ ,
\ee
where the tensors with little-group indices $a,b,c,..$ are symmetric. The parenthesis denote total symmetrization of the little-group indices, which includes a $1/n!$ factor. The multiplet has eight bosonic and eight fermionic degrees of freedom. $\psi^\alpha_{abc}$ are the gravitini and the vector $A^{a b}$ represents the graviphoton. 

The above graviton multiplet can be obtained as a double copy of the ${\cal N}=2$ vector multiplet and a non-supersymmetric vector field,
\be
{\cal H}^{{\cal N}=2}_{abc} = {\cal V}^{{\cal N}=2}_{(a} \otimes  A_{bc)} \ ,
\ee
where the little-group indices are symmetrized. If instead we antisymmetrize a pair of indices (contracting with $\epsilon_{ab}$), we get the axidilaton-vector multiplet 
\be
{\cal V}^{{\cal N}=2}_{c} = {1 \over 2} \epsilon^{ab} {\cal V}^{{\cal N}=2}_{a} \otimes   A_{bc} = \chi^1_c+ \Big(A_{c b}^{\rm axi}+ \, {\epsilon_{cb}\over \sqrt{2}} \varphi^{\rm dil} \Big) \eta^b+   \chi^2_c \, (\eta)^2 \ .
\ee

Next considering ${\cal N}=4$, we find the maximally-supersymmetric vector multiplet
\be
{\cal V}^{{\cal N}=4} = \phi + \chi^\alpha_a \eta^a_\alpha+  A_{a b} \eta^a_\alpha \eta^{b\alpha}+ \phi^{\alpha \beta} \eta_{a\alpha} \eta_{ \beta}^a+\bar \chi^\alpha_a \eta^a_\alpha (\eta)^2+ \bar \phi (\eta)^4 \ , \label{N4Vsuperfield} 
\ee
where $(\eta)^2= \frac{1}{4}\eta^a_\alpha  \eta_a^\alpha$. It contains eight bosonic and eight fermionic degrees of freedom, and it has manifest $SU(2)\times U(1)\subset USp(4)$ R-symmetry. The vector multiplet can also be obtained from two different double copies:
\be
{\cal V}^{{\cal N}=4} = {1 \over 2}\Phi_{{\cal N}=2}^\alpha \otimes \Phi_{{{\cal N}=2}} ^\beta \epsilon_{\alpha \beta}
= {1 \over 2} {\cal V}^{{\cal N}=2}_{\dot a} \otimes {\cal V}^{{\cal N}=2}_{\dot b} \epsilon^{\dot a \dot b} \ .
\ee

The ${\cal N}=4$ graviton multiplet transforms as a little-group vector,
\bea
{\cal H}^{{\cal N}=4}_{ab} & =& A_{a b}+ \Big(\psi^\alpha_{abc} + {1\over \sqrt{2}} \psi^\alpha_{(a} \epsilon{\vphantom{\psi^\alpha}}_{b)c} \Big)\eta^c_\alpha+  \Big(h_{abcd}+ {1\over \sqrt{2}} \tilde A_{(a(c}\epsilon_{b)d)} + {1\over 2} \epsilon_{a(c}\epsilon_{d)b} \phi\Big)\eta^c_\alpha \eta^{d \alpha} \nn \\ 
&& \null  + \eta_{c\alpha} \eta^{c}_\beta  A_{a b}^{\alpha\beta} + \Big(\bar \psi^\alpha_{abc} + {1\over \sqrt{2}} \bar \psi^\alpha_{(a} \epsilon\vphantom{\psi^\alpha}_{b)c}\Big)\eta^c_\alpha (\eta)^2+ \bar A_{a b}(\eta)^4 \ ,
\eea
where the six graviphotons are distributed as three singlets and a triplet of the manifest $SU(2)$ R symmetry. The multiplet has 24 bosonic and 24 fermionic degrees of freedom. It can be obtained as the double copy of an ${\cal N}=4$ vector and non-supersymmetric vector,
\be
{\cal H}^{{\cal N}=4}_{ab} = {\cal V}^{{\cal N}=4} \otimes  A_{ab} \ ,
\ee
or, alternatively, as the symmetric double copy of two ${\cal N}=2$ vector multiplets:
\be
{\cal H}^{{\cal N}=4}_{ab} = {\cal V}^{{\cal N}=2}_{(a} \otimes {\cal V}^{{\cal N}=2}_{b)} \ .
\ee
The little-group indices are symmetrized, and the single-copy Grassmann variables $\eta^{\alpha}_a =(\eta_a, \tilde\eta_a )$ form a doublet of the expected manifest $SU(2)$ R symmetry of the double-copy multiplet.

Finally, the ${\cal N}=6$ and maximal ${\cal N}=8$ supersymmetry graviton multiplets can be obtained as the double copies
\be
{\cal H}^{{\cal N}=6}_{a} = {\cal V}^{{\cal N}=4} \otimes {\cal V}^{{\cal N}=2}_{a}\,, ~~~~~~~{\cal H}^{{\cal N}=8} = {\cal V}^{{\cal N}=4} \otimes {\cal V}^{{\cal N}=4} \ ,
\ee
respectively. It is not difficult to carry out the multiplication and separate the various monomials in Grassmann variables to identify the component states. We will not write them out explicitly, but instead note that these multiplets contain $64+64$ and $128+128$ bosonic and fermionic states, respectively. 

\subsection{On-shell superfields for massive 5D multiplets}

From the perspective of the little group, 5D massive multiplets and massless 6D multiplets are the same. We will therefore use the established notation for the latter to also denote the former.

The 5D massive vector multiplet can also be seen as a 6D chiral $(1,0)$ multiplet,\footnote{We label the massive supermultiplets in terms of the $(r_+,r_-)$ $SU(2) \times SU(2)$ little-group components of the supercharge~\cite{Hull:2000cf}.}
\be
{\cal V}^{(1,0)}_{\dot a} = \chi_{\dot a}^{1}+ W_{a \dot a} \eta^a + \chi_{\dot a}^{2} (\eta)^2 \ ,
\ee
where the fermions $\chi^{\dot a}_{\alpha}$ are symplectic-Majorana spinors. Similarly, the hypermultiplet is
\be
\Phi_{(1,0)}^{\dot \alpha} =  \varphi^{1 \dot \alpha} + \chi_a^{\dot \alpha} \eta^a+ \varphi^{2 \dot  \alpha} \, (\eta)^2 \ .
\ee
The above fermions $\chi_{\dot a}^{\alpha}$ and scalars $\varphi^{\alpha \dot \alpha}$ are transforming in the $SO(4)$ R-symmetry, but the realization of the supersymmetry algebra using $\eta^a$ partially breaks the R-symmetry to $ U(1) \times SU(2)$. 

An $(1,0)$ tensor  multiplet can be written as
\be
{\cal T}^{(1,0)}_{a} = \chi_{a}^{1}+ \Big(B_{a b }{+}  \, {\epsilon_{ab} \over \sqrt{2}} \varphi \Big) \eta^b + \chi_{a}^{2} \, (\eta)^2\,,
\ee
where $B_{a b }$ is a self-dual tensor,  and the fermions $\chi_{a}^{\alpha}$ are symplectic-Majorana spinors of opposite $SO(4)$ chirality compared to the spinors in the $(1,0)$ vector multiplet.  

A 6D $(1,0)$ graviton multiplet, and the corresponding massive 5D multiplet,~is
\be
{\cal H}^{(1,0)}_{a \dot a \dot b} =\psi_{a \dot a \dot b}^{1} + \Big(h_{ab \dot a \dot b} + {\epsilon_{ab} \over \sqrt{2}}B_{\dot a \dot b}  \Big) \eta^b+ \psi_{a \dot a \dot b}^{2} (\eta)^2 \ .
\ee
It has 12 bosonic and 12 fermionic states and it is symmetric in the dotted little-group indices. $B_{\dot a \dot b}$ is the anti-self-dual gravitensor and $\psi_{a \dot a \dot b}^{\alpha}$ are the gravitini.
Through the double copy, we can also write the $(1,0)$ graviton multiplet as
\be
{\cal H}^{(1,0)}_{a \dot a \dot b} = {\cal V}^{(1,0)}_{\dot a}  \otimes W_{a \dot b } - \epsilon_{\dot a \dot b}\,{\cal T}^{(1,0)}_{a} \ ,
\label{2copygraviton}
\ee
where we ``subtract" the tensor multiplet following the ghost prescription outlined in Ref.~\cite{Johansson:2014zca}. 
The subtracted tensor multiplet has its own double-copy decomposition,
\be
{\cal T}^{(1,0)}_{a} = {1 \over 2} \epsilon_{\dot\alpha \dot\beta}\Phi_{(1,0)}^{\dot  \alpha} \otimes \chi_{a}^{\dot  \beta}\, .
\ee
The construction \eqref{2copygraviton} corresponds to the double-copy realization of pure 6D $(1,0)$ supergravity via tensor ghosts from Ref.~\cite{Johansson:2014zca}.  
It should be noted that we can also identify ${\cal T}^{(1,0)}_{a}= {\cal V}^{(1,0)}_{\dot a}  \otimes W_{a \dot b } \epsilon^{\dot a \dot b}$.

It is interesting to contrast the double-copy form of some of the component fields in Eq.~\eqref{2copygraviton}.  The gravitensor is 
\be
B_{\dot a \dot b} = \frac{1}{2} W_{a (\dot a} \otimes   \tilde W_{b \dot b)} \, \epsilon^{ab} \ ,
\ee
while the ``axitensor'' contained in the first product and subtracted by ${\cal T}^{(1,0)}_{a}$ are
\be
\widetilde B_{a b}^{\rm axi} = -\frac{1}{2}  W_{(\vphantom{b}a \dot a} \otimes   \tilde W_{b) \dot b} \, \epsilon^{\dot a \dot b} = \frac{1}{2}\chi_{a}^\alpha \otimes \chi_{b \alpha} \ .
\ee
The second identification is necessary for the subtraction in Eq.~\eqref{2copygraviton} to remove the ``axitensor'' together with its partners and yield the (1,0) graviton multiplet.

We turn next to maximal supersymmetry. The $(1,1)$ vector multiplet (or, equivalently, the massive 5D ${\cal N}=4$ vector multiplet) is
\bea
{\cal V}^{(1,1)} &=&\varphi_{1\dot 1} +  \chi_{a\dot 1} \eta^a+  \chi_{1 \dot a} \etaT^{\dot a}  +\varphi_{1\dot 2}  (\etaT)^2 +\varphi_{2\dot 1}  (\eta)^2   
\nn \\&&\null
+ W_{a \dot a} \eta^a   \etaT^{\dot a}    + \chi_{a \dot 2} \eta^a (\etaT)^2  + \chi_{2 \dot a} \etaT^{\dot a} (\eta)^2 +\varphi_{2\dot 2}  (\eta)^2  (\etaT)^2 \ .
\eea
Introducing the variables $\zeta^{\alpha}= \big(1, (\eta)^2\big)$ and  $\tilde \zeta^{\dot \alpha}= \big(1, (\etaT)^2\big)$, it can also be written more compactly as
\be
{\cal V}^{(1,1)} =  \varphi_{\alpha \dot \alpha} \zeta^{\alpha} \tilde\zeta^{\dot \alpha} + \chi_{a \dot \alpha} \eta^a \tilde\zeta^{\dot \alpha} + \chi_{\alpha \dot a} \etaT^{\dot a} \zeta^{\alpha}  + W_{a \dot a} \eta^a   \etaT^{\dot a} \ .
\ee
If we assign $SO(4)\sim SU(2)\times SU(2)$ transformations to the $\zeta,\tilde\zeta$ variables, the massive ${\cal N}=4$ vector multiplet  exhibits the complete $SO(4)$ R-symmetry unbroken by the central charge.

The $(2,0)$ tensor multiplet can be written as
\be
{\cal T}^{(2,0)} =  \phi + \chi^\alpha_a \eta^a_\alpha+  B_{a b} \eta^a_\alpha \eta^{b\alpha}+ \phi^{\alpha \beta} \eta_{a\alpha} \eta_{ \beta}^a+\bar \chi^\alpha_a \eta^a_\alpha (\eta)^2+ \bar \phi (\eta)^4 \ ,
\ee
where states exhibit a manifest $SU(2)\times U(1) \subset SO(5)$ R-symmetry. This multiplet has exactly the same form as the massless ${\cal N}=4$ vector multiplet, except that the massless vector $A_{a b}$ is here replaced by a massive self-dual tensor $B_{a b}$. The  $(2,0)$ tensor multiplet can be obtained as a double copy in two different ways,
\bea
{\cal T}^{(2,0)} &=& {1 \over 2}\Phi_{(1,0)}^{\dot \alpha} \otimes \Phi_{{(1,0)}}^{\dot \beta} \epsilon_{\dot\alpha \dot\beta} = {1 \over 2} {\cal V}^{(1,0)}_{\dot a} \otimes {\cal V}^{(1,0)}_{\dot b} \epsilon^{\dot a \dot b} \ ,
\eea
in terms of two $(1,0)$ hypermultiplets or two $(1,0)$ vector multiplets, respectively.

The $(1,1)$ graviton multiplet can be written as a double copy,
\be
{\cal H}^{(1,1)}_{b \dot b}  = {\cal V}^{(1,1)} \otimes W_{b \dot b}  =  W_{\alpha \dot \alpha b \dot b} \zeta^{\alpha}\tilde\zeta^{\dot \alpha} +\tilde \psi_{a \dot \alpha b \dot b} \eta^a \tilde\zeta^{\dot \alpha} + \tilde \psi_{\alpha \dot a b \dot b} \etaT^{\dot a} \zeta^{\alpha}  + \tilde h_{a \dot a b \dot b} \eta^a   \etaT^{\dot a} \ ,
\ee
where $W_{\alpha \dot \alpha b \dot b}$ are four massive vectors, $\tilde \psi_{a \dot \alpha b \dot b} =  \psi_{a \dot \alpha b \dot b} + \epsilon_{ab}\psi_{ \dot \alpha \dot b}/\sqrt{2} $ consist of a massive gravitino and a fermion (similarly for $\tilde \psi_{\alpha \dot a b \dot b}$), and $\tilde  h_{a \dot a b \dot b} = h_{a \dot a b \dot b}+ \epsilon_{ab} B_{ \dot a \dot b}/\sqrt{2}  + \epsilon_{\dot a \dot b} B_{a b}/\sqrt{2} + \epsilon_{ab}  \epsilon_{\dot a \dot b} \phi/2 $ consists of the graviton together with a tensor and a scalar.  

The $(2,0)$ graviton multiplet can also be obtained as a double copy,
\be
{\cal H}^{(2,0)}_{\dot a \dot b} =  {\cal V}^{(1,0)}_{\dot a} \otimes {\cal V}^{(1,0)}_{\dot b} -  \epsilon_{\dot a \dot b} \, {\cal T}^{(2,0)}\, ,
\ee
as the little-group-traceless part of the product of two $(1,0)$ vector multiplets. 

In a similar spirit, $(2,0)$ non-metric graviton multiplet is a double copy of 
a $(2,0)$ tensor multiplet and another tensor,
\be
{\cal H}^{(2,0)}_{a b} =  {\cal T}^{(2,0)} \otimes B_{ a b} \ .
\ee
The ``graviton'' $h_{abcd}$ is non-metric in the sense that it only has chiral little-group indices and descends from the field strength of a mixed tensor gauge field and not that of a symmetric metric tensor in six dimensions. It contains five degrees of freedom (i.e. the same as the 5D massless graviton). 

For 3/4-maximal and maximal supersymmetry, we can obtain massive graviton multiplets as various double copies. The most interesting are the following:
\bea
&& {\cal H}^{(2,1)}_{\dot a} = {\cal V}^{(1,1)} \otimes {\cal V}^{(1,0)}_{\dot  a}\,, ~~~~~~~~~~~~~~{\cal H}^{(2,2)} = {\cal V}^{(1,1)} \otimes  {\cal V}^{(1,1)} \,, \nn \\
&& {\cal H}^{(3,0)}_{a} = {\cal T}^{(2,0)} \otimes  {\cal T}^{(1,0)}_{a}\,, ~~~~~~~~~~~~~~{\cal H}^{(4,0)} = {\cal T}^{(2,0)} \otimes   {\cal T}^{(2,0)} \ .
\eea
We will not spell out here the details of the component fields, as they are simply a straightforward but tedious exercise of distributing the multiplets over the tensor product and identifying fields. 
  
In the massive case, we can also consider gravitino multiplets of Poincar\'{e} supergravity. They appear when supersymmetry is spontaneously broken, and can be obtained from a double-copy construction. When maximal supersymmetry is partially broken down to 3/4-maximal supersymmetry, the massive 5D gravitino multiplet is of $(2,1)$ type,
\be
\Psi_{\dot \alpha}^{(2,1)} = {\cal V}^{(1,1)} \otimes \Phi^{(1,0)}_{\dot \alpha} \ .
\ee
For 3/4-maximal supersymmetry partially broken down to 1/2-maximal supersymmetry, the gravitino multiplet is of $(1,1)$ type 
\be
\Psi_{\dot a \alpha}^{(1,1)} = {\cal V}^{(1,1)} \otimes \chi_{\dot a \alpha} = {\cal V}^{(1,0)}_{\dot a} \otimes  \Phi^{(0,1)}_{\alpha} \ ,
\ee
or $(2,0)$ type
\be
\Psi_{\dot a \dot\alpha}^{(2,0)} = {\cal V}^{(1,0)}_{\dot a} \otimes  \Phi^{(1,0)}_{\dot \alpha} \ .
\ee
Finally, for partial breakings that preserve 1/4-maximal supersymmetry, the gravitino multiplets are of $(1,0)$ type
\be
\Psi_{\dot a \dot b \alpha}^{(1,0)} = {\cal V}^{(1,0)}_{(\dot a} \otimes \chi_{\dot b) \alpha} \ .
\ee

\section{Five-dimensional amplitudes with vectors and tensors\label{sec4}}

In our conventions, $m$-point color-dressed gauge-theory amplitudes are written as
\begin{equation}
\mathbb{A}_m = g^{m-2} \sum_{\sigma \in S_{m-2}} A_m (1, \sigma(2), \cdots ,\sigma(m-1),m) {\rm Tr} (T^{1} T^{\sigma(2)} \cdots T^{\sigma(m-1)} T^{m}) \ ,
\end{equation}
where $A_m(1,\ldots,m)$ are color-ordered partial amplitudes and representation matrices are chosen to obey ${\rm Tr} (T^a T^b) = \delta^{ab}$ and $[T^a, T^b] = \tilde f^{abc} T^c$. We use ${\cal A}_m(1,\ldots,m)$ to denote the corresponding partial superamplitudes. Similarly, we use $M_m$ to denote component gravitational amplitudes and ${\cal M}_m$ to denote gravitational superamplitudes.

\subsection{Three-point amplitudes and superamplitudes \label{3pointAmps}}

As illustrated in Section \ref{sec2}, we dress the $SU(2)$ little-group indices with auxiliary bosonic variables $\auxu_a,\auxv_{\dot a}$. To avoid explicitly displaying little-group indices, we use the short-hand notation introduced in Eq.~(\ref{compactnotation}),
\be
|\,\ket{i\,} \equiv | k^a_i \rangle \auxu_{ia}  
\,,~~~
|\ket{q_i} \equiv | q^a_i \rangle \auxu_{ia}  
\,,~~~
|\,\aket{i\,} \equiv | {\bm p}^a_i \rangle \auxu_{ia}
\, ,~~~
|\,\sket{i\,} \equiv | {\bm p}^{\dot a}_i ] \auxv_{i{\dot a}} \ .
\ee
For example, in this notation, the massless polarization vector corresponding to the $i$-th leg is written as
\begin{equation}  
\varepsilon^\mu_i =  {\bra{\,i}\,| \Gmat^\mu | \ket{\qref{i}}  \over \sqrt{2}}  \ ,
\end{equation}
where $q_i$ is a reference momentum obeying the condition (\ref{q_constraint}), as well as  $\langle q_i q_j \rangle=0$.  

The three-gluon partial amplitude in 5D Yang-Mills theory is given by
\begin{equation}
A_3(1A,2A,3A) = {i} \big( \auxu_{2}^a \langle 2_a {\qref{3}}_b \rangle \auxu^b_3 \big)\big( \auxu_{3}^c\langle  3_c {\qref{2}}_d \rangle \auxu^d_2 \big) \big( \auxu_{1}^e\langle  1_e 2_f \rangle \langle  2^f {\qref{1}}_g \rangle \auxu^g_1 \big) + \text{cyclic}(1,2,3) \ ,
\end{equation}
where we have chosen to display the little-group indices and to collect little-group singlets in parenthesis. Using the short-hand notation, this expression can be rewritten in the more compact form
\begin{equation}
\label{3A}
A_3(1A,2A,3A)  
= {i }   \bra{2 } \ket{\qref{3}}  \bra{3} \ket{\qref{2}}     \bra{1} |k_2| \ket{\qref{1}}  + \text{cyclic}(1,2,3) \ .
\end{equation}
Three-point partial amplitudes in 5D Yang-Mills theory with massless matter are
\begin{eqnarray}
\label{phi2A}
A_3(1\phi,2\phi,3A) &=& 
{ i  }  \bra{3} 1_a \rangle \langle  1^a \ket{\qref{3}} = {  i  } \bra{3} |k_1| \ket{\qref{3}} \ ,  \\
A_3(1 \chi^2,2 \chi^1, 3 A) &=& { i  }   \bra{1}\ket{3}    \bra{\qref{3}} \ket{2}  - (3 \leftrightarrow q_3) \ ,\label{ampf}\\
\label{chi2phi}
A_3(1 \chi^2,2  \chi^1, 3 \phi) &=&  -{ i \over \sqrt{2} }   \bra{1}\ket{2} \ .
\end{eqnarray}
These amplitudes can be obtained from  
the superamplitude
\be
{\cal A}_{3}^{\cN=2} (1{\cal V},2{\cal V},3{\cal V}) =  -i  \eta_{1}^a \langle {1}_a|p_2| \ket{\qref{1}}   \Big( \sum_{i=1}^3 \eta_{i}^b \langle {i}_b\ket{\qref{2}}\Big) \Big( \sum_{j=1}^3 \eta_{j}^c \langle j_c \ket{  \qref{3}}  \Big) + \text{cyclic}(1,2,3)\,, \label{super2} 
\ee
by acting with the appropriate Grassmann derivatives corresponding to the desired external states. More explicitly, the component amplitudes in Eqs.~\eqref{3A}-\eqref{chi2phi} are
\bea
A_3(1A,2A,3A) &=& -\Big( \auxu_1^a{\partial \over \partial \eta^a_1} \Big)
\Big( \auxu_2^b{\partial \over \partial \eta^b_2} \Big)
\Big( \auxu_3^c{\partial \over \partial \eta^c_3} \Big)
{\cal A}_{3}^{\cN=2} (1{\cal V},2{\cal V},3{\cal V}) \ , \\
A_3(1\phi,2\phi,3A) &=& 
- \Big( {\epsilon^{ab}\over \sqrt{2}} {\partial \over \partial \eta^a_1} {\partial \over \partial \auxu^b_1}  \Big)
\Big( {\epsilon^{cd} \over \sqrt{2}} {\partial \over \partial \eta^c_2} {\partial \over \partial \auxu^d_2}  \Big)
\Big( \auxu_3^e{\partial \over \partial \eta^e_3} \Big)
{\cal A}_{3}^{\cN=2} (1{\cal V},2{\cal V},3{\cal V})  , \qquad \quad \\
A_3(1 \chi^2,2\chi^1,3A) &=&  \Big({\partial \over \partial \eta^2_1} {\partial \over \partial \eta^1_1} \Big)
\Big( \auxu_3^c{\partial \over \partial \eta^c_3} \Big)
{\cal A}_{3}^{\cN=2} (1{\cal V},2{\cal V},3{\cal V}) \ , \\
A_3(1 \chi^2,2\chi^1,3 \phi) &=&  \Big( {\partial \over \partial \eta^2_1} {\partial \over \partial \eta^1_1} \Big)
\Big( {\epsilon^{cd} \over \sqrt{2}} {\partial \over \partial \eta^c_3} {\partial \over \partial \auxu^d_3}  \Big)
{\cal A}_{3}^{\cN=2} (1{\cal V},2{\cal V},3{\cal V}) \ , 
\eea
where  the operators corresponding to each leg have been constructed so that they extract the components of the corresponding on-shell superfields with the correct normalization.

The superamplitude (\ref{super2}) can also be rewritten in terms of the supercharges $|Q\rangle =\sum_i |\qi_i \rangle = \sum_i |i^a \rangle \eta_{ia}$ as 
\begin{eqnarray}
{\cal A}_{3}^{\cN=2} (1{\cal V},2{\cal V},3{\cal V}) &=& - i  \langle  \qi_1 | p_2 | \ket{\qref{1}}  \ \bra{\qref{2}} | Q \rangle
\bra{\qref{3}} | Q \rangle  + \text{cyclic}(1,2,3)\,. \label{super2b}
\end{eqnarray}
Note that this object is totally symmetric under permutations. This is appropriate given the fermionic nature of the superfields since the  color factor (which has been stripped off) is totally antisymmetric. 	In a similar way, we can write down the superamplitude between two massless hypermultiplets and one vector multiplet in 5D $\cN=2$ super-Yang-Mills theory. It has an even simpler expression,
\begin{eqnarray}
{\cal A}_{3}^{\cN=2} (1{\Phi}^{\dot \alpha},2{\Phi}^{\dot\beta},3{\cal V}) &=& {i \over 2} \epsilon^{\dot \alpha \dot \beta}  \bra{  \qref{3}}  | Q \rangle  \langle Q | Q \rangle  \label{super2hyper} \ .
\end{eqnarray}

Moving to the massive case, it is instructive to first consider three-point amplitudes between two massive spinors and a massless vector. With a chiral spinor $\chi$ and an anti-chiral spinor $\tilde \chi$, we have the following candidate amplitudes:
\bea
A(1\chi,2\chi, 3A) &=& -  {i  \sqrt{2}}  \abra{1} |  \varepsilon_3 | \aket{2} =   {i} \abra{1}\ket{3} \bra{\qref{3}}\aket{2}  - (3 \leftrightarrow q_3 ) \ , \\
A(1\tilde \chi,2\tilde \chi, 3A) &=&
-  {i \sqrt{2}}  \sbra{1} |  \varepsilon_3 | \sket{2} =  i \sbra{1}\ket{3} \bra{\qref{3}}\sket{2}  - (3 \leftrightarrow q_3 ) \ , \\
A(1 \chi, 2\tilde \chi, 3A) &\stackrel{?}{=}& - {i  \sqrt{2}} \abra{1} |  \varepsilon_3 | \sket{2} =  i \abra{1}\ket{3} \bra{\qref{3}}\sket{2}  - (3 \leftrightarrow q_3 )  \ .
\eea
Taking the two masses to have opposite sign, i.e. $m_1=m=-m_2$, one can check that the third candidate amplitude $A(1 \chi, 2\tilde \chi, 3A)$ is not gauge invariant. For $\varepsilon_3 \rightarrow p_3$ this amplitude does not vanish, unlike the other two, and hence it must vanish identically. This implies that $\chi$ and $\tilde \chi$ are not related by CPT symmetry, and hence massive 5D theories have a notion of chirality that is preserved by the interactions, similar to 6D massless theories.  

We also give the component amplitude between two massive and one massless vectors, which has the following expression:
\begin{eqnarray}
A_3(1W,2W,3A) &=& -{i \over 4 m^2} \Big\{     \big( \sbra{2} \ket{3}  \bra{\qref{3}} \aket{2} - (3 \leftrightarrow q_3) \big) \sbra{1} |k_3| \aket{1}  \\
&& \qquad  
\null +{1 \over 2}\big( \abra{1}\aket{2} \sbra{1}\sket{2} - \abra{1}\sket{2} \sbra{1}\aket{2}  ) \bra{3}  |p_1| \ket{\qref{3}} \Big\} - (1 \leftrightarrow 2)\,.  \no
\end{eqnarray}
It corresponds to the partial amplitude in Yang-Mills theory with spontaneously-broken gauge symmetry (hence the $W$-boson label) or, alternatively, to a Kaluza-Klein Yang-Mills theory.

If we have three massive vector bosons, the spontaneously-broken 5D Yang-Mills amplitude can be cast in the following form,\footnote{Note that the identity
$
(m_1 - m_3)\sbra{1}\sket{2} \abra{2}\aket{3} \sbra{3}\aket{1} \pm  \text{perms} =0 
$
can be used to rewrite the amplitude.}
\begin{equation}
A_3(1W,2W,3W) =  { i \over 4 m_2m_3}  \sbra{1}\sket{2} \abra{2}\aket{3} \sbra{3}\aket{1} \pm  \text{perms}(1,2,3) \ , \label{sampN2mass}
\end{equation}
where the permutations run over the dihedral group $S_3$ with negative sign for odd permutations. The amplitude we give here is only supported on the mass conservation relation,
\be
m_1 + m_2 + m_3 = 0 \ . 
\label{massconservation}
\ee
It can be derived from the superamplitude of 5D half-maximal super-Yang-Mills theory on the Coulomb branch,
\bea
{\cal A}_{3}^{(1,0)} (1{\cal V},2{\cal V},3{\cal V}) &=&  {i \over 24} {(m_1-m_2) \sbra{1} \sket{2} \sbra{3}Q\rangle \over m_1m_2m_3} \langle Q | Q \rangle  +  \text{cyclic}(1,2,3) \ , 
\eea
which is invariant under chiral (1,0) supersymmetry, provided that the mass conservation condition in Eq. (\ref{massconservation}) is obeyed. The bilinears in the chiral and anti-chiral supercharges can be written more explicitly as 
\begin{equation}
 \langle Q | Q \rangle  = \sum_{i,j=1}^3 \eta_{i}^a \langle \bm i_a \bm j_b \rangle \eta_j^b  \ ,
 \qquad 
  [ \tilde Q | \tilde  Q ]  = \sum_{i,j=1}^3 \tilde \eta_{i}^{\dot a} [ \bm i_{\dot a} \bm j_{\dot b} ] \tilde \eta_j^{\dot b}  \ .
\end{equation}

The (1,0) superamplitude for two massive hypermultiplets and a massive Coulomb-branch vector multiplet in 5D super-Yang-Mills theory is
\bea
{\cal A}_{3}^{(1,0)} (1\Phi^{\dot \alpha},2\Phi^{\dot\beta},3{\cal V}) &=&
  i {\epsilon^{\dot \alpha \dot \beta} \over 4m_3} \sbra{\bm 3}Q\rangle  \langle Q | Q \rangle\,,
\eea
where we have again assumed mass conservation. 

Moving to the case of massless $\cN=4$ supersymmetry, we find no straightforward generalization of the superamplitude (\ref{super2}). 
To explore other possible forms, let us turn to the known 6D construction of three-point amplitudes~\cite{Cheung:2009dc,Dennen:2009vk}, and adapt it to 5D. We begin by noticing that, with massless three-point kinematics, all the 2-by-2 matrices ${\langle i^a j^b \rangle}$ have rank one, and hence have no inverses. Instead one can decompose them into $SU(2)$ little-group spinors $u_{i}^a$ as 
\begin{eqnarray}
\langle 1^a 2^b \rangle = u_{1}^a u_2^b, & \qquad &  \langle 2^a 1^b \rangle = -u_{1}^b u_2^a, \\
\langle 2^a 3^b \rangle = u_{2}^a u_3^b, &&  \langle 3^a 2^b \rangle = -u_{2}^b u_3^a, \\
\langle 3^a 1^b \rangle = u_{3}^a u_1^b, &&  \langle 1^a 3^b \rangle = -u_{3}^b u_1^a\,.
\end{eqnarray}
All  $u_{i}^a$ variables are uniquely determined, up to an overall  sign, by this system. 

There exist corresponding reference spinors $w_i^a$ that satisfy 
\begin{equation}
  u^a_i w^b_i -u^b_i w^a_i = \epsilon^{ab}  ~~~~~~\text{(no sum over $i$)}
\end{equation}
and that can be further constrained by
\begin{equation} \label{mom_cons1}
  \sum_{i=1}^3  | i_a\rangle w_i^a =0 \ .
\end{equation}
With these constraints, $w_i^a$ are not unique, but the remaining two degrees of freedom will cancel out once the amplitude is assembled~\cite{Cheung:2009dc}. Note that the $u_i^a$ and $w_i^a$ variables only exists for on-shell three-point kinematics, which has degenerate and complex momenta.

With the above variables, the massless three-point superamplitude in 5D $\cN=4$ super-Yang-Mills theory can be written as 
\begin{eqnarray} \label{super4}
{\cal A}_{3}^{\cN=4} (1{\cal V},2{\cal V},3{\cal V}) &=& {i \over 4} \delta^2 \big( \sum_{i} \eta^{\alpha}_{i a} w^a_i \big)
\prod_{\alpha=1,2} \langle Q^\alpha | Q^\alpha \rangle \ ,
\end{eqnarray}
where $|Q^\alpha \rangle = \sum_i |Q_i^\alpha \rangle= \sum_i |i^a \rangle\eta^\alpha_{ia}$ is the supercharge. Note that one cannot write down a delta function of all the eight supercharges as this object vanishes for three-point kinematics.\footnote{The spinors $| Q_i^\alpha \rangle$, with $i=1,2,3$, span a three-dimensional subspace of $USp(2,2)$ and  $\delta^{8}(Q)$ has the interpretation as the square of the corresponding four-volume, hence it vanishes.}
Nevertheless,  is possible to verify that the above superamplitude is invariant under all supersymmetry generators. 

So far, we have only given implicit definitions for the $u^a_i$ and $w^a_i$ spinors, but we can do a bit better. Given that our spinor parametrization $|i^a\rangle$ is linear in momentum in the first little-group component $a=1$, we have via momentum conservation the convenient relation
\be \label{mom_cons2}
\sum_{i=1}^3 |i^{1}\rangle = 0\,.
\ee
This implies that $\langle 1^1 2^1  \rangle = \langle 2^1 3^1  \rangle= \langle 3^1 1^1  \rangle \equiv r^2$, which then gives the unique solution for the $u^a_i$ spinors,
\be
u_1^a = \frac{1}{r}\langle 1^a 2^1\rangle\,,~~\qquad ~u_2^a = \frac{1}{r}\langle 2^a 3^1\rangle\,,~\qquad~~u_3^a = \frac{1}{r}\langle 3^a 1^1\rangle \ .
\ee
One can easily confirm that $u_i^a u_j^b$  gives by construction three correct entries of the corresponding 2-by-2 matrix $\pm \langle i^a j^b \rangle$, and by the reduced rank it then follows that the forth entry ($a=b=2$) is also correct. Our choice of parametrization also gives a simple solution for the $w^a_i$ spinors,  
\be
w^a_i = w^a = \frac{1}{r} (0,1)\,.
\ee
With this choice, \eqn{mom_cons1} follows from \eqn{mom_cons2}, while $u_i^{[a} w_{\vphantom{i}}^{b]}=\epsilon^{ab}$ follows from the fact that all little-group spinors have the same first entry $u_i^{1}=r$. 

We may further simplify the construction by noticing that, for any choice of the $w_i^a$ variables, one can find a corresponding global reference spinor $\langle \rho|$ such that
\be
w_i^a =\langle \rho| i^a \rangle \,.
\ee
Thus, we can write also the last delta function in \eqn{super4} in terms of the supercharge,
\be
{\cal A}_{3}^{\cN=4} (1{\cal V},2{\cal V},3{\cal V}) = {i \over 4} 
 \prod_{\alpha=1,2} \langle \rho| Q^\alpha \rangle  \langle Q^\alpha | Q^\alpha \rangle ={i \over 36} \prod_{\alpha=1,2} \det\big( |Q^\alpha\rangle,|Q^\alpha \rangle,|Q^\alpha \rangle,|\rho\rangle\big)\,.
\ee
Because the last formula is a 4-by-4 determinant, there exists a three-fold family of $|\rho\rangle$ spinors that gives the same amplitude. For the choice of $w_i^a=\frac{1}{r} (0,1)$, we find that the global reference spinor can be chosen simply as
\be
\langle \rho| = \frac{1}{r}  (0,0,0,1)\,.
\ee

Generalization to the massive case follows even more closely the 6D case~\cite{Cheung:2009dc}. Three point kinematics implies that
\begin{equation}
\det \langle {\bm i^a \bm j^b} \rangle = m_i m_j = \det [ {\bm i^{\dot a} \bm j^{\dot b}} ] \ , \qquad
\det \langle {\bm i^a \bm j^{\dot a}} ] = 0 \ .
\end{equation}
One can define the little-group spinors
\begin{eqnarray}
\langle \bm 1^a \bm  2^{\dot b} ] = u_{1}^a \tilde u_2^{\dot b}, & \qquad \qquad &  \langle \bm 2^a \bm 1^{\dot b} ] = - u_2^a \tilde u_{1}^{\dot b} \ , \\
\langle \bm 2^a \bm 3^{\dot b} ] = u_{2}^a \tilde u_3^{\dot b}, &&  \langle \bm 3^a  \bm 2^{\dot b} ] = - u_3^a \tilde u_{2}^{\dot b} \ , \\
\langle \bm 3^a \bm 1^{\dot b} ] = u_{3}^a \tilde u_1^{\dot b}, &&  \langle \bm 1^a \bm 3^{\dot b} ] = - u_1^a \tilde u_{3}^{\dot b} \ ,
\end{eqnarray}
with corresponding reference spinors $w_i^a$ and $\tilde w_i^{\dot b}$ that satisfy
\begin{eqnarray}
  u^a_i w^b_i -u^b_i w^a_i &=& \epsilon^{ab}  ~~~\text{(no sum over $i$)} \ , \\
  \tilde u^{\dot a}_i  \tilde w^{\dot b}_i - \tilde u^{\dot b}_i \tilde w^{\dot a}_i &=& \epsilon^{{\dot a} \dot b}  ~~~\text{(no sum over $i$)}   \ ,
\end{eqnarray}
with the additional constraints
\begin{equation}
  \sum_i  | \bm i\rangle_a w_i^a =0 \ , \qquad  \ \sum_i  |\bm i]_{\dot b} \tilde w_i^{\dot b} =0\,.
\end{equation}
With our parametrization of the massive spinors, an explicit solution is
\be
u_i^a = \frac{1}{r}\langle {\bm i}^a {\bm j}^1]\,,~~~~~  \tilde u_i^{\dot a} =  \frac{1}{r}[ {\bm i}^{\dot a} {\bm j}^1\rangle\,,~~~~~ w_i^a=\frac{1}{r} (0,1)=  \tilde w_i^{\dot a}\,,
\ee
where ${\bm j}= {\bm i}+1$~Mod~3, and $r^2=\langle {\bm 1}^1 {\bm 2}^1]=\langle {\bm 2}^1 {\bm 3}^1]=\langle {\bm 3}^1 {\bm 1}^1]$.

The three-point massive $(1,1)$ superamplitude in maximal 5D super-Yang-Mills on the Coulomb branch is then the direct generalization of \eqn{super4},\footnote{Recall the vector amplitude from Ref.~\cite{Cheung:2009dc}, $(u_1^a u_2^b w_3^c +{\rm cyclic}(1,2,3)) (\tilde u_1^{\dot  a}  \tilde u_2^{\dot  b} \tilde w_3^{\dot c} +{\rm cyclic}(1,2,3))$, may be obtained from \eqn{super4massive} using the identities $\langle {\bm 1}^a {\bm 2}^b \rangle = u^a_1 u^b_2 - 2 m_2 w_1^a u_2^b  + 2 m_1 u_1^a w_2^b $ and 
$\langle {\bm 1}^a {\bm 2}^b \rangle w_3^c -u_1^a u_2^b w_3^c +{\rm cyclic}(1,2,3)=0$.} 
\begin{eqnarray}
{\cal A}_{3}^{(1,1)} (1{\cal V},2{\cal V},3{\cal V}) &=& {i \over 4 } \delta \big( \sum_{i} \eta^{}_{i a} w^a_i \big) \delta \big( \sum_{i} \tilde \eta^{}_{i \dot a} \tilde w^{\dot a}_i \big)  \langle Q | Q \rangle  [ \tilde Q | \tilde Q ] \nn \\
&=& {i \over 36 } \det\big(| Q \rangle,| Q \rangle,| Q \rangle,| \rho \rangle \big)\, \det\big(| \tilde Q ],| \tilde Q ],| \tilde Q ],| \rho  ] \big)\,,
\label{super4massive}
\end{eqnarray}
where the global reference spinors satisfy $\langle \rho| i^a \rangle = w_i^a$ and $[ \rho| i^{\dot a} ]= \tilde w_i^{\dot a}$, which for our simple choice can be obtained by $\langle \rho|= [ \rho|= \frac{1}{r}(0,0,0,1)$. Again, we have imposed mass conservation.

It is interesting to note that, from the simple factorized form of the massive $(1,1)$ superamplitude, one may attempt to write down a massive $(2,0)$ superamplitude for non-abelian self-dual tensor multiplets (interpreted as a Kaluza-Klein reduction of a 6D $(2,0)$ tensor theory). The naive guess is
\be
{\cal A}_{3}^{(2,0)} (1{\cal T},2{\cal T},3{\cal T}) \stackrel{?}{=} {i \over 4 }  \prod_{\alpha=1,2} \langle \rho | Q^\alpha \rangle   \langle Q^\alpha | Q^\alpha \rangle =  {i \over 36 } \prod_{\alpha=1,2} \det\big(| Q^\alpha \rangle,| Q^\alpha \rangle,| Q^\alpha \rangle,| \rho \rangle \big)\,,
\label{super20massive}
\ee
which superficially looks indistinguishable from the massless ${\cal N}=4$ superamplitude. However, we do not find any covariant expressions (in terms of momenta and tensor polarizations) that match the corresponding candidate non-abelian three-tensor component amplitude
\be
A_{3}(1B^{ab},2B^{cd},3B^{ef}) \stackrel{?}{=}   i\det\big(| {\bm 1}^{(a} \rangle,| {\bm 2}^{(c} \rangle,| {\bm 3}^{(e} \rangle,| \rho \rangle \big) \det\big(| {\bm 1}^{b)} \rangle,| {\bm 2}^{d)} \rangle,| {\bm 3}^{f)} \rangle,| \rho \rangle \big)\,,
\ee
nor for the corresponding candidate scalar-tensor component amplitude
\be
A_{3}(1\phi,2\phi,3B^{ab}) \stackrel{?}{=}   i  u_3^a u_3^b = i\sqrt{2}m_3 \varepsilon_{3,\mu \nu}^{ab} \frac{[ {\bm 1}^1|\gamma^{\mu \nu}| {\bm 2}^1 ]}{[{\bm 1}^1 {\bm 2}^1]}\,.
\ee
The fact that we do not find covariant formulas is consistent with the three-point-amplitude analysis of Ref.~\cite{Czech:2011dk}, and indicate that these are not well-behaved amplitudes. In the next subsection the corresponding four-tensor amplitude will be analyzed with the same conclusion. 

The three-graviton superamplitude in massive 5D Kaluza-Klein (2,0) supergravity is given by the double copy
\begin{eqnarray}
{\cal M}_{3}^{(2,0)} (1{\cal H},2{\cal H},3{\cal H}) &=&-i{\cal A}_{3}^{(1,0)} (1{\cal V},2{\cal V},3{\cal V})  {\cal A}_{3}^{(1,0)} (1{\cal V},2{\cal V},3{\cal V}) \\
&=&  {i \over 36 }  \det\big(| {\bm 1} ],| {\bm 2}],| {\bm 3} ],| \rho ] \big)^2 \prod_{\alpha=1,2} \det\big(| Q^\alpha \rangle,| Q^\alpha \rangle,| Q^\alpha \rangle,| \rho \rangle \big)\,. \nn 
\label{superHmassive}
\end{eqnarray}
Likewise, there exist a well-behaved massive abelian tensor-graviton superamplitude in the Kaluza-Klein (2,0) supergravity theory
\begin{eqnarray}
&&{\cal M}_{3}^{(2,0)} (1{\cal T},2{\cal T},3{\cal H}^{\dot a \dot b}) =- i {\cal A}_{3}^{(1,0)} (1{\cal V}^{\dot c},2{\cal V}^{\dot d},3{\cal V}^{\dot a})  {\cal A}_{3}^{(1,0)} (1{\cal V}_{\dot c},2{\cal V}_{\dot d},3{\cal V}^{\dot b}) \nn \\
&&=  {i \over 36 }  \det\big(| {\bm 1}^{\dot c} ],| {\bm 2}^{\dot d}],| {\bm 3}^{\dot a} ],| \rho ] \big)\det\big(| {\bm 1}_{\dot c} ],| {\bm 2}_{\dot d}],| {\bm 3}^{\dot b} ],| \rho ] \big) \prod_{\alpha=1,2} \det\big(| Q^\alpha \rangle,| Q^\alpha \rangle,| Q^\alpha \rangle,| \rho \rangle \big)\, \nn \\
&&= -{i \over 18 }  \tilde u_3^{\dot a} \tilde u_3^{\dot b}  \prod_{\alpha=1,2} \det\big(| Q^\alpha \rangle,| Q^\alpha \rangle,| Q^\alpha \rangle,| \rho \rangle \big)\,.
\label{superTTHmassive}
\end{eqnarray}

Note the same tensor amplitude can be obtained as the double copy of $(1,0)$ super-Yang-Mills amplitudes with massive half-hyper multiplets and a Coulomb-branch vector, \be
{\cal M}_{3}^{(2,0)} (1{\cal T},2{\cal T},3{\cal H}^{\dot a \dot b})
=
-i {\cal A}_{3}^{(1,0)} (1 \Phi,2\Phi,3{\cal V}^{\dot a})  {\cal A}_{3}^{(1,0)} (1\Phi,2\Phi,3{\cal V}^{\dot b})
\,,
\ee
and the equivalence of the two double copies follow from the kinematic identities 
\begin{eqnarray}
\tilde u_3^{\dot a} \tilde u_3^{\dot b}&=&-\frac{1}{2}\det\big(| {\bm 1}^{\dot c} ],| {\bm 2}^{\dot d}],| {\bm 3}^{\dot a} ],| \rho ] \big)\det\big(| {\bm 1}_{\dot c} ],| {\bm 2}_{\dot d}],| {\bm 3}^{\dot b} ],| \rho ] \big) \nn \\
&=& 
-\frac{1}{4}\det\big(| {\bm 1}_{\dot c} ],| {\bm 1}^{\dot c}],| {\bm 3}^{\dot a} ],| \rho ] \big)\det\big(| {\bm 2}_{\dot d} ],| {\bm 2}^{\dot d}],| {\bm 3}^{\dot b} ],| \rho ] \big)\nn \\
\tilde u_3^{\dot a} &=& \frac{1}{2}\det\big(| {\bm 1}_{\dot c} ],| {\bm 1}^{\dot c}],| {\bm 3}^{\dot a} ],| \rho ] \big)=-\frac{1}{2}\det\big(| {\bm 2}_{\dot d} ],| {\bm 2}^{\dot d}],| {\bm 3}^{\dot a} ],| \rho ] \big)\,,
\end{eqnarray}
where, as before,  mass conservation is assumed. The above equivalence agrees with the fact that the different tensors can be embedded into the $(2,2)$ gravitational theory where they are related by R-symmetry.

The three-graviton superamplitude in massive Kaluza-Klein (2,2) supergravity~is
\begin{eqnarray}
{\cal M}_{3}^{(2,2)} (1{\cal H},2{\cal H},3{\cal H}) &=&- i {\cal A}_{3}^{(1,1)} (1{\cal V},2{\cal V},3{\cal V})  {\cal A}_{3}^{(1,1)} (1{\cal V},2{\cal V},3{\cal V}) \\
=&& \!\!\! \!\!\!\! {i \over 36^2 }  \prod_{\alpha=1,2} \det\big(| Q^\alpha \rangle,| Q^\alpha \rangle,| Q^\alpha \rangle,| \rho \rangle \big) \!
\prod_{\dot \alpha=1,2}
\det\big(| \tilde Q^{\dot \alpha} ],| \tilde Q^{\dot \alpha} ],| \tilde Q^{\dot \alpha} ],| \rho ] \big)\,, \nn 
\end{eqnarray}
and similarly in (1,1) supergravity 
\begin{eqnarray}
{\cal M}_{3}^{(1,1)} (1{\cal H},2{\cal H},3{\cal H}) &=&- i {\cal A}_{3}^{(1,0)} (1{\cal V},2{\cal V},3{\cal V}) {\cal A}_{3}^{(0,1)} (1{\cal V},2{\cal V},3{\cal V}) \nn \\
&=&  {i \over 36 }  \det\big(| {\bm 1} \rangle,| {\bm 2}\rangle,| {\bm 3} \rangle,| \rho \rangle \big) \det\big(| Q \rangle,| Q \rangle,| Q \rangle,| \rho \rangle \big)  \\
&& ~ \times \, 
\det\big(| {\bm 1} ],| {\bm 2}],| {\bm 3} ],| \rho ] \big) \det\big(| \tilde Q ],| \tilde Q ],| \tilde Q ],| \rho ] \big)\,, \nn
\end{eqnarray}
where we assumed that both massive amplitudes originate from massless 6D kinematics via Kaluza-Klein compactification. All the massive amplitudes given above can equivalently describe 6D massless amplitudes.

\subsection{Superamplitudes at four points}

The most convenient expressions for superamplitudes are relevant for gauge theories with maximal $\cN=4$ supersymmetry. As it is to be expected, the basic object appearing at four points is 
the Grassmann delta function, which takes the following expression with the 5D spinor-helicity notation,
\begin{equation}
\delta^8 (Q) ={\frac{1}{4}}\prod_{\alpha=1}^2 \big(\sum_{i < j} \eta_{i}^{\alpha a} \langle {i_a j_b } \rangle \eta_j^{\alpha b} \big)^2 ={\frac{1}{64}}\prod_{\alpha=1}^2 \big( \langle Q^{\alpha} | Q^{\alpha}\rangle \big)^2 \ .
\end{equation}
The color-ordered superamplitude between four vector multiplets is then readily written as
\begin{equation}
{\cal A}_{4}^{\cN=4}(1 {\cal V},2 {\cal V},3 {\cal V},4 {\cal V}) =- 4 i {\delta^8 (Q)  \over st} \ .
\end{equation}
We can directly verify that this superamplitude yields the appropriate component amplitudes. For example, the amplitude between four massless vectors is
\begin{eqnarray}
A_4(1 A,2A,3A,4A) \! &=&  -{i \over st} \big(\bra{1}\ket{2} \bra{3}\ket{4}+ \text{cyclic}(1,2,3) \big)^2 \nn \\
&=&  {i \over st} \Big\{
2 \bra{1}\ket{2} \bra{2}\ket{3} \bra{3}\ket{4} \bra{4}\ket{1} -
\bra{1}\ket{2}^2 \bra{3}\ket{4}^2
+ \text{cyclic}(1,2,3) \Big\}  . \qquad
\label{amp4vec}
\end{eqnarray}

The $\cN=2$ vector superfields can then be embedded in the $\cN=4$ superfields, which leads to the identification 
\begin{eqnarray}
{\cal A}_{4}^{\cN=2}(1 {\cal V},2{\cal V}, 3{\cal V}, 4{\cal V}) & = & {-}4i \Big( z_1^a {\partial \over \partial \eta_1^{a2} }\Big)\! \cdots \! \Big( z_4^d {\partial \over \partial \eta_4^{d2} }\Big) {\delta^8 (Q)  \over st} \no \\
& = & -2i{\delta^4 (Q)  \over st} \big(\bra{1}\ket{2}\bra{3}\ket{4}+{\rm cyclic}(1,2,3)\big) .
\end{eqnarray}
Since the complement of an $\cN=2$ vector superfield in an $\cN=4$ vector superfield is a hypermultiplet, from the amplitude above one can extrapolate that the amplitudes with hypermultiplets take a similar form,
\be
{\cal A}_{4}^{\cN=2}(1 \Phi^{\dot \alpha},2\Phi^{\dot \beta}, 3{\cal V}, 4{\cal V}) = -2i \epsilon^{\dot \alpha \dot\beta} {\delta^4 (Q)  \over st} \bra{3}| p_1|\ket{4}\ ,
\ee
and
\be
{\cal A}_{4}^{\cN=2}(1 \Phi^{\dot \alpha},2\Phi^{\dot \beta}, 3\Phi^{\dot \gamma}, 4\Phi^{\dot \delta})  = -2i\delta^4 (Q) \Big({\epsilon^{\dot \alpha \dot\beta}\epsilon^{\dot \gamma \dot\delta} \over s}+{\epsilon^{ \dot\beta \dot \gamma}\epsilon^{ \dot\delta \dot \alpha}  \over t}\Big)\ .
\ee

It is easy to promote the massless four-point amplitudes to massive amplitudes by replacing the expressions for the massless supercharges with the supercharges in the massive case. As before, we have several distinct representations of the supersymmetry algebra that differ based on the chirality of the supercharges under the $SO(4)$ little group. Let us start by introducing the needed Grassmann delta functions for massive supercharges. For standard non-chiral $(1,1)$ supersymmetry, we have
	\be
	\delta^4 (Q)\delta^4 (\tilde Q) = {1 \over 4} \big(\sum_{i<j} \eta_{i}^{a} \langle { \bm i_a \bm j_b } \rangle \eta_j^{ b} \big)^2 \big(\sum_{i<j} \tilde \eta_{i}^{\dot a} [ { \bm i_{\dot a} \bm j_{\dot b} } ] \tilde \eta_j^{ \dot b} \big)^2 = {1 \over 64}  \big(\langle Q | Q\rangle [ \tilde Q | \tilde Q] \big)^2\,,
    \ee
whereas for chiral (2,0) and (0,2) supersymmetry one has the following Grassmann delta functions:
	\begin{eqnarray}
	\delta^8 (Q) &=& {1 \over 4} \prod_{\alpha=1}^2 \big(\sum_{i<j} \eta_{i}^{\alpha a } \langle { \bm i_a \bm j_b } \rangle \eta_j^{\alpha b} \big)^2 = {1 \over 64} \prod_{\alpha=1}^2 \big(\langle Q^\alpha | Q^\alpha\rangle \big)^2 \ , \qquad \\
	\delta^8 (\tilde Q) &=&  {1 \over 4} \prod_{\alpha=1}^2 \big(\sum_{i,j} \tilde \eta_{i }^{\alpha \dot a} [ { \bm i_{\dot a} \bm j_{\dot b} } ] \tilde \eta_j^{\alpha \dot b} \big)^2  = {1 \over 64} \prod_{\alpha=1}^2 \big([ \tilde Q^\alpha | \tilde Q^\alpha]\big)^2 \nn \ .
	\end{eqnarray}
Based on these expressions, we can obtain massive superamplitudes between four massive vector multiplets,
\begin{equation}
{\cal A}_{4}^{(1,1)}(1{\cal V},2{\cal V},3{\cal V},4{\cal V}) = -4i {\delta^4 (Q) \delta^4 (\tilde Q)  \over (s-m_{s}^2 )( t-m_{t}^2)} \ ,
\end{equation}
where $m_s,m_t$ denote the masses for the $s$- and $t$-channel poles. They are given by the mass-conservation condition: $m_s=m_1+m_2$, $m_t=m_2+m_3$. 
It is useful to explicitly write out the component amplitude between four massive vector fields, it is
\begin{eqnarray}
A_4(1W,2W,3W,4W) \! &=& 
-i {\big(\abra{1}\aket{2} \abra{3}\aket{4}+ \text{cyclic}(1,2,3)\big)\big(\sbra{1}\sket{2} \sbra{3}\sket{4}+ \text{cyclic}(1,2,3)\big) \over (s-m_s^2)(t-m_t^2)}  \nn \\
&=& 
\! {i \over (s-m_s^2)(t-m_t^2)} \Big\{
\abra{1}\aket{2} \sbra{2}\sket{3} \abra{3}\aket{4} \sbra{4}\sket{1} +
\sbra{1}\sket{2} \abra{2}\aket{3} \sbra{3}\sket{4} \abra{4}\aket{1} -  \no \\
&& \qquad  \qquad \qquad \sbra{1}\sket{2} \abra{2}\aket{1} \sbra{3}\sket{4} \abra{4}\aket{3}
+ \text{cyclic}(1,2,3) \Big\}  . \qquad
\end{eqnarray}

Having obtained the $(1,1)$ vector amplitude, one may guess that a naive candidate for the color-ordered superamplitude between four massive tensor $(2,0)$ multiplets is
\begin{equation}
{\cal A}_{4}^{(2,0)}(1{\cal T},2{\cal T},3 {\cal T},4 {\cal T}) \stackrel{?}{=} - 4i {\delta^8 (Q)  \over (s-m_s^2)(t-m_t^2)} \ .
\end{equation}
To check if this is a well-behaved amplitude we want to check if it factorizes properly. In terms of the already guessed three-point amplitudes, we can work out the needed $s$-channel factorization,
\be
\int d^4\eta_5 {\cal A}_{3}^{(2,0)}(1{\cal T},2{\cal T},5 {\cal T}) {\cal A}_{3}^{(2,0)}(-5{\cal T},3{\cal T},4 {\cal T}) = - 4i \delta^8 (Q)  {\tau^2 \over (t-m_t^2)^2}\,,
\ee
where $\tau= \frac{1}{r_{34}} \big([{\bm 1}^1 {\bm 2}^1] [{\bm 3}^1 \rho] +[{\bm 2}^1 {\bm 3}^1] [{\bm 1}^1 \rho]+[{\bm 3}^1 {\bm 1}^1] [{\bm 2}^1 \rho]\big)$ and $\rho= \frac{1}{r_{12}}(0,0,0,1)$, with $r^2_{ij}=\langle {\bm i}^1 {\bm j}^1]$. In order for the factorization to work, we need the double pole to cancel out. However, the relation between $\tau$ and the pole is $\tau \tilde \tau= t-m_t^2$, where  $\tilde \tau$ is given by swapping square and angle spinors in $\tau$. Hence, the double pole only cancels out in the massless limit or, alternatively, if we had done the same calculation for $(1,1)$ super-Yang-Mills theory.  One may wonder if the three-point amplitudes can be modified such that they absorb the unwanted $\tau^2/(t-m_t^2)$ factor. However, given that it  depends non-trivially on momenta belonging to different three-point amplitudes, this is unlikely to work out. Our conclusion agree with the analysis of Ref.~\cite{Cachazo:2018hqa}, where the non-abelian (2,0) candidate amplitude was shown to have irreconcilable factorization properties. 

In contrast to the problematic non-abelian tensor case, there is interesting and well-behaved gravitational amplitude that involves abelian tensor multiplets.  The four-tensor superamplitude in (2,0) supergravity is given by the double copy
\bea
{\cal M}_{4}^{(2,0)}\!(1{\cal T},2{\cal T},3 {\cal T},4 {\cal T}) \! &=& \!- i (s \! - \! m_s^2)  {\cal A}_{4}^{(1,0)}\!(1{\cal V}\!,2{\cal V}\!,3{\cal V}\!,4{\cal V}) {\cal A}_{4}^{(1,0)}\!(1{\cal V}\!,2{\cal V}\!,4{\cal V}\!,3{\cal V}) \nn  \\
&=& i \delta^8 (Q) \Big( { 1 \over s-m_s^2}+{ 1 \over t-m_t^2}+{1  \over u-m_u^2} \Big) \ ,
\eea
where the poles corresponds to (massive) graviton exchange.

\subsection{Comparison with superamplitudes in other dimensions}

Massive multiplets in five dimensions are closely related with massless multiplets in six dimensions since the little group is the same. This relation was used in \cite{Czech:2011dk} to dimensionally-reduce the 6D spinor-helicity formalism to 5D and discuss certain three-point amplitudes potentially related to those of the (2,0) 6D theory.

Following the standard 6D notation, a massless momentum is written as~\cite{Gunaydin:1999ci,Dennen:2009vk,Cheung:2009dc}
\begin{equation}
p^{AB}_{[6]} = \lambda^{Aa} \lambda^B_a\,,~~~~~~~~p_{AB}^{[6]} = \tilde \lambda_{A \dot a} \tilde\lambda_B^{\dot a}\,, 
\end{equation}
where $\lambda^A_a$ are 6D chiral spinors, $A,B=1,\ldots,4$, and $a,b$ are $SU(2)$ indices, and similarly for the anti-chiral spinors $\tilde \lambda^A_{\dot a}$. In our massive formalism, we have the natural identification
\begin{equation}
\lambda^A_a  \equiv \langle \bm p_a |^A\,,  \qquad 
\tilde \lambda_A^{\dot a} \equiv |\bm p^{\dot a}]_A\,.
\end{equation}
The sixth component of the 6D momentum is interpreted as the mass in five dimensions (leading to a natural interpretation for the mass-conservation condition $\sum_i m_i=0$). 
Formulas for the polarization vectors and supercharges follow straightforwardly from their 6D analogues. It should also be noted that the $USp(2,2)$-invariant matrix $\Omega$ can be seen as the sixth entry to the $\Sigma$ matrices in 6D, where the remaining five entries are given by the 5D gamma matrices.  

Moving to the massless case, it can be obtained from the 6D spinor-helicity formalism through the identification of the chiral and anti-chiral spinors $\lambda$ and $\tilde \lambda$.
In a sense, massless superfields and amplitudes in five dimensions present a closer analogy with  4D superamplitudes in a non-chiral representation \cite{Huang:2011um}. Starting from the conventional chiral representation of four-point amplitudes, one can act with a Grassmann Fourier transform with respect to half of the Grassmann coordinates,
\begin{equation}
\Phi=\int d\eta^3 d\eta^4 e^{\eta^3 \tilde\eta^3+ \eta^4 \tilde\eta^4} \Phi_\text{chiral} \ .
\end{equation}
For example, the maximally supersymmetric on-shell vector superfield in four dimensions now have the expansion
\begin{eqnarray}
{\cal V}_{\rm 4D}^{\cN=4} &=&  \varphi^{34} + \bar \chi_{\alpha 34} \eta^\alpha +  \chi_{{\tilde \alpha}}  \etaT^{\tilde \alpha} + A^{-} \eta^2 + A^{+}  \etaT^2 + \varphi_{\alpha {\tilde \alpha}} \eta^{\alpha}  \etaT^{\tilde \alpha} + \no \\ && \qquad \qquad \chi_{\alpha}  \eta^\alpha  (\etaT)^2  + \bar \chi_{12 {\tilde \alpha}}   \etaT^{\tilde \alpha} (\eta)^2 + \varphi^{12} (\eta)^2  (\etaT)^2 ,
\end{eqnarray}
with $\alpha=1,2$ and $ {\tilde \alpha}=3,4$. This closely reflects our 5D superfields.
The corresponding superamplitude at four points in 4D spinor-helicity notation is given by
\begin{equation}
{\cal A}_{\rm 4D}^{\cN=4}(1,2,3,4) = -i { \prod_{\alpha=1}^2 \big(\sum_{i<j} \eta_i^\alpha \langle { i j } \rangle \eta_j^\alpha)  \ \prod_{{\tilde \alpha}=3}^4 \big(\sum_{i<j}  \etaT_i^{\tilde \alpha} [ { i j }]  \etaT_j^{\tilde \alpha} )  \over st} \ ,
\end{equation} 
The relation between 4D and 5D variables is then simply
\begin{equation}
\eta_{ia}^{\alpha} \langle { i^a j_b } \rangle \eta_j^{b\alpha} \Big|_{\text{4D kin}} =
\eta_i^\alpha \langle { i j } \rangle \eta_j^\alpha + \tilde \eta_i^{\tilde \alpha} [{ i j }] \tilde \eta_j^{\tilde \alpha} \ \ .
\end{equation}
It is also immediate to verify that, with this relation, Eq. (\ref{amp4vec}) reproduces the familiar expression for the 4D amplitudes upon dimensional reduction with the correct normalization.

\subsection{Amplitude decomposition in terms of total $SU(2)$ weights}

We now turn to the massless amplitudes in five dimensions, and ask what is the 5D equivalent of the 4D decomposition in helicity sectors. Since the 4D helicity follows the weight under little group rephrasing and the 4D little group is a subgroup of the 5D one, $U(1)\subset SU(2)$, it is natural to expect that the desired decomposition 
is that in irreducible representations of the 5D little group.

With the particular choice of reference momenta 
from Eq.~(\ref{choiceq}), we have the relations
\begin{equation}
(\varepsilon_i^{11} \cdot \varepsilon_j^{11} ) = 0 \ , \qquad
(\varepsilon_i^{11} \cdot \varepsilon_j^{12} ) = 0 \ , \qquad
(\varepsilon_i^{22} \cdot \varepsilon_j^{12} ) = 0 \ , \qquad
(\varepsilon_i^{22} \cdot \varepsilon_j^{22} ) = 0 \ , \label{epsilonzero}
\end{equation}  
for all $i,j$. A three-point massless amplitude can be decomposed in terms of the total $SU(2)$ little-group representation carried by the external states. This is done by tensoring three three-dimensional representations of $SU(2)$,
\begin{equation}
{\bf 3 } \otimes {\bf 3 } \otimes {\bf 3 }  = {\bf 7 } \oplus 2 \times {\bf 5 } \oplus 3 \times {\bf 3 } \oplus {\bf 1} \ .
\end{equation}
 To generate all amplitudes within a given $SU(2)$ sector, we can start from the lowest-weight state in which there is a maximal number of $1$ in the little-group labels for the polarization, and act with raising operators. Eq. (\ref{epsilonzero}) has as a consequence that we need at least two $1$ and two $2$ labels to have a nonzero results, hence the $\bf {7}$ and ${\bf 5}$ sectors are automatically zero for YM theory in 5D, since the corresponding lowest-weight states do not have this property. 
 This therefore constrains the structure of the allowed polynomials in auxiliary
 variables that are dressing the amplitudes with free $SU(2)$ indices:
 \begin{equation}
 A^{(3)}_{a_1 b_1 \ldots a_3 b_3} \auxu_1^{a_1}\auxu_1^{b_1}\auxu_2^{a_2}\auxu_2^{b_2}\auxu_3^{a_3}\auxu_3^{b_3} =
 A^{(3)}_{a_1 b_1 \ldots a_3 b_3} \Big\{ \auxu_1^{a_1}\auxu_1^{b_1}\auxu_2^{a_2}\auxu_2^{b_2}\auxu_3^{a_3}\auxu_3^{b_3} \Big|_{\bm 3} + 
\auxu_1^{a_1}\auxu_1^{b_1}\auxu_2^{a_2}\auxu_2^{b_2}\auxu_3^{a_3}\auxu_3^{b_3} \Big|_{\bm 1} \Big\} \ ,
 \end{equation}
 where we have defined
 $\auxu_1^{a_1}\auxu_1^{b_1}\auxu_2^{a_2}\auxu_2^{b_2}\auxu_3^{a_3}\auxu_3^{b_3} \Big|_{\bm 3} $ and  $\auxu_1^{a_1}\auxu_1^{b_1}\auxu_2^{a_2}\auxu_2^{b_2}\auxu_3^{a_3}\auxu_3^{b_3} \Big|_{\bm 1} $ as the degree-six polynomials in the auxiliary variables that transform in the three-dimensional $SU(2)$ representation and in the singlet representation.
 
At four points, the decomposition is carried out in a similar way except that now the starting point is the product of four three-dimensional representations of $SU(2)$:
\begin{equation}
{\bf 3 } \otimes {\bf 3 } \otimes {\bf 3 } \otimes {\bf 3 }  = {\bf 9 } \oplus 3 \times {\bf 7 } \oplus 6 \times {\bf 5 } \oplus 6 \times {\bf 3 } \oplus 3 \times {\bf 1} \ .
\end{equation}
As before, the amplitudes in the ${\bm 9}$ and ${\bm 7}$ sectors are equal to zero, mirroring the vanishing of all plus and one-minus amplitudes in four dimensions. As in four dimensions, these sectors become nonzero once a higher-dimensional operator is added to the Lagrangian of the theory.

\section{More on double-copy amplitudes \label{sec5}}

In this section, we revisit the double-copy construction of Maxwell-Einstein and Yang-Mills-Einstein theories with the new formalism.

\subsection{Maxwell-Einstein and Yang-Mills-Einstein supergravities revisited}

 The main advantage of the 5D formalism is that it does not require reducing the supergravity Lagrangian to four dimensions to match amplitudes from the double copy, as it was done in the earlier work~\cite{Chiodaroli:2014xia,Chiodaroli:2015wal}. Being able to formulate the construction directly in five dimensions streamlines and simplifies the derivation. 
One of the remarkable properties of Maxwell-Einstein theories with $\cN=2$ supersymmetry in five dimensions is that they exhibit the simple cubic vector couplings 
\cite{Gunaydin:1983bi,Gunaydin:1984nt}
\begin{equation}
{1 \over 6 \sqrt{6}}\int d^5x \, C_{IJK} \epsilon^{\mu \nu \rho \sigma \tau} F^I_{\mu \nu} F^J_{\rho \sigma} A^K_{\tau}  \ ,
\end{equation}
where the $C_{IJK}$ tensor is symmetric in the $I,J,K$ indices running over the number of vector fields present in the theory.
The full Lagrangian of the theory is fixed once the $C_{IJK}$ tensor in determined, as done in Refs. \cite{Gunaydin:1983bi,Gunaydin:1984ak,Gunaydin:1986fg}. In turn, this tensor can be read off the three-vector amplitude, which has the following simple expression,
\be
M_3(1A,2A,3A) \! = \! -i  { \sqrt{ 8 \over 3}}  C_{IJK} \epsilon \big(k_1,\varepsilon_1,k_2, \varepsilon_2, \varepsilon_3 \big)  
\! = \! - i  {2 C_{IJK}  \over  \sqrt{3}} \bra{1}\ket{2} \big( \bra{1}\ket{3}  \, \bra{\qref{3}}\ket{2} 
- (3 \! \leftrightarrow \! q_3)  \big) \ , \label{Camp}
\ee
where $\epsilon(a,b,c,d,f)= a^{\mu}b^{\nu}c^{\rho}d^{\sigma}f^{\lambda}\epsilon_{\mu\nu \rho \sigma \lambda}$.
In short, these theories are completely specified by their three-point amplitudes.

The simplest double-copy realization of a family of such theories involves, as one of the gauge theory factors, $\cN=2$ super Yang-Mills theory and, as the other factor, a Yang-Mills theory with additional adjoint scalars $\phi^i$. At three points, all $\cN=2$ amplitudes can be packaged in a single superamplitude. The two non-zero amplitudes on the non-supersymmetric side are $A_3(1A,2A,3A)$ and $A_3(1A,2\phi^i,3\phi^j)$. With the intent of identifying the relevant $C_{IJK}$ tensors, we focus on amplitudes between vectors in the supergravity theory. In five dimensions, we have three possible double-copy origins for vector states: 
\begin{eqnarray}
A_{ab}^{i} &=& A_{ab} \big|_{\cN=2} \otimes \phi^i \big|_{\cN=0} \ , \\
A_{ab}^{1} &=& \phi \big|_{\cN=2} \otimes A_{ab} \big|_{\cN=0} \ , \\
A_{ab}^{0} &=&  A_{c(a} \big|_{\cN=2} \otimes A^{c}_{b)}\big|_{\cN=0} \ , 
\end{eqnarray}
where the gauge-theory scalars carry an index $i=2,\ldots, n_V$ and we label the vector states with their little-group indices. It is not difficult to see that there are only two non-zero amplitude between three vectors: $M_3 (1A^0,2A^i,3A^j)$ and $M_3 (1A^0,2A^1,3A^1)$. The former is given by\footnote{Here the normalization of the double-copy follows the KLT formula with the gravitational coupling set to $\kappa=2$, i.e. $M_3 =  -{i} A(1,2,3) \tilde A(1,2,3)$.}
\begin{equation}
M_3 (1A^0,2A^i,3A^j) = -  {i \over 4}  { \epsilon^{ab} } {\partial \over \partial \auxu_1^{a}} A_3 (1A,2A,3A)  {\partial \over \partial \auxu_1^{b}} A_3 (1A,2\phi^i,3\phi^j) \ ,
\end{equation}
where the differential operators in the antisymmetric indices extract the correct little-group representation. After some additional work, it is possible to cast the double-copy amplitude in the form 
\begin{equation}
M_3 (1A^0,2A^i,3A^j) = 
  {i}  {\delta^{ij} } \bra{1}\ket{2} \big( \bra{1}\ket{3}  \, \bra{\qref{3}}\ket{2} 
- (3 \leftrightarrow q_3)  \big) \ .
\label{0ij}
\end{equation}
The second nonzero amplitude is 
\begin{eqnarray}
M_3 (1A^0,2A^1,3A^1) &=& - i    {\epsilon^{ab} \over 4} {\partial \over \partial \auxu_1^{a}} A_3 (1A,2\phi,3\phi)  {\partial \over \partial \auxu_1^{b}} A_3 (1A,2A,3A) \no \\
&=& - i \bra{1}\ket{2} \big( \bra{1}\ket{3}  \, \bra{\qref{3}}\ket{2} 
- (3 \leftrightarrow q_3)  \big)  \ ,
\label{011}
\end{eqnarray}
which despite its appearance is invariant under the full $S_3$ permutation symmetry. The difference in sign compared to Eq.~\eqref{0ij} can be traced to the fact that in the two amplitudes the scalars entering the double-copy construction for the supergravity vectors come from different gauge-theory factors.  

We can now read off the $C_{IJK}$ for this theory directly by comparing the double-copy amplitudes in Eqs.~\eqref{011} and \eqref{0ij} with the supergravity expression Eq.~(\ref{Camp}):
\begin{equation}
C_{011} =  { \sqrt{3}  \over \ 2}  \ , \qquad C_{0ij} = - {\sqrt{3 }  \ \over 2}  \delta^{ij} \ .
 \end{equation}
This is the well-known generic Jordan family of Maxwell-Einstein supergravities in five dimensions \cite{Gunaydin:1983bi,Gunaydin:1984ak,Gunaydin:1986fg}. It should be noted that analyzing amplitudes in five dimensions gives us the option to avoid explicitly expanding around a base point.\footnote{There is however the built in assumption that there exists a base point which leads to canonically-normalized vector fields. It is this point that is chosen by the double copy.} The earlier formulation of the double-copy construction for these theories requires matching of amplitudes in four dimensions, so the 5D spinor-helicity formalism considerably streamlines the derivation.

We can extend the supersymmetric gauge theory by including hypermultiplets and the non-supersymmetric theory by including fermions. This yields extra vectors in the double-copy theory as 
\begin{equation}
A_{ab}^{\alpha} = \chi^{\vphantom{\alpha}}_{(a} \big|_{\cN=2} \otimes \chi^\alpha_{b)} \big|_{\cN=0} \ , \\
\end{equation}
where $\alpha$ is an extra flavor index carried by the gauge-theory fermions. Introducing these fields turns on additional three-vector amplitudes, which are of the form
\begin{equation}
M_3 (1A^i,2A^\alpha,3A^\beta) = - i   A_3 (1A,2\chi,3\chi)  A_3 (1 \phi^i,2 \chi^\alpha,3 \chi^\beta) \ .
\end{equation}
The amplitude between two fermions and one scalar in the non-supersymmetric theory is taken to be proportional to a matrix in the global indices that appears in the Yukawa couplings,
\begin{equation}
A_3 (1 \phi^i,2 \chi^\alpha,3 \chi^\beta) = - {i \over \sqrt{2}} \Gamma^i_{\alpha\beta} \bra{2} \ket{3} \ .
\end{equation}
The resulting supergravity amplitude is
\begin{equation}
M_3 (1A^i,2A^\alpha,3A^\beta) = -  {i \over  \sqrt{2}}  \Gamma^{i}_{\alpha \beta} \bra{2}\ket{3} \big( \bra{2}\ket{1}  \ \bra{\qref{1}}\ket{3} 
- (3 \leftrightarrow q_3)  \big)  \ ,
\end{equation}
which leads to the additional non-zero entries of the $C_{IJK}$ tensor,
\begin{equation}
C_{i \alpha \beta} = {1 \over 2}\sqrt{3\over 2}  \Gamma^i_{\alpha \beta} \ .
\end{equation}
We have recovered  the construction for homogeneous supergravities first given in Ref.~\cite{Chiodaroli:2015wal}. Here  we do not discuss the constraints imposed by color/kinematics duality on the matrix $\Gamma^i_{\alpha \beta}$, i.e. we just assume that we have gauge theories obeying color/kinematics duality from which we can take the amplitudes entering the double copy. Color/kinematics duality can be conveniently studied at the level of the gauge theories, and in this particular case requires that 
the $\Gamma^i_{\alpha \beta}$ matrices obey Clifford-algebra relations, as shown in Ref.~\cite{Chiodaroli:2015wal}. The net result is that we recover the classification of homogeneous Maxwell-Einstein supergravities from the supergravity literature \cite{deWit:1991nm}.

For completeness, we also give the double-copy map for the other bosonic supergravity fields,\footnote{
Note that the field combinations appearing in the double-copy map are canonically normalized, with the exception of $\varphi^1$, which has a ${3 \over 2}(\partial_\mu \varphi^1)^2$ kinetic term in the supergravity Lagrangian expanded at the appropriate base-point.}
\begin{align}
h_{abcd} &= A_{(ab} \big|_{\cN=2} \otimes A_{cd)} \big|_{\cN=0} \ , \\
\varphi^{1} &=  A_{ab} \big|_{\cN=2} \otimes A^{ab}\big|_{\cN=0} \ , \\
\varphi^{i} &= \phi \big|_{\cN=2} \otimes \phi^i \big|_{\cN=0} \ , \\
\varphi^{\alpha} &= {1 \over \sqrt{2}} \chi_a \big|_{\cN=2} \otimes \chi^{a \alpha} \big|_{\cN=0} \ . 
\end{align}
Finally, it is completely straightforward to turn on trilinear scalar couplings in the non-supersymmetric gauge theory entering the construction while preserving the duality between color and kinematics. This yields non-Abelian gauge interactions in the gravitational double-copy theory, which becomes a Yang-Mills-Einstein theory with gauge symmetry given by the flavor symmetry of the trilinear couplings in the non-supersymmetric gauge theory Lagrangian, as shown in Ref.~\cite{Chiodaroli:2014xia}.

The component amplitudes discussed in this section can be  supersymmetrized using the superamplitudes introduced in Section~\ref{sec4}. 
The amplitude between three supergravity massless vector multiplets can be written as\footnote{Here, the indices $i,j,k$ run over the matter vector multiplets in the theory. This expression requires the $C_{IJK}$ tensors to be given in the canonical basis with $C_{000}=1$, $C_{0ij}=-{1 \over 2} \delta^{ij}$, $C_{00i}=0$ and $C_{ijk}$ arbitrary.} 
\begin{eqnarray}
{\cal M}_3 (1 {\cal V}^i,2 {\cal V}^j,3 {\cal V}^k)  &=& -i {2 \over \sqrt{3}} C_{ijk} \bra{1}\ket{2}  \bra{  \qref{3}}  | Q \rangle \langle Q | Q \rangle    \ , \label{MESGTsuper}
\end{eqnarray}
where we used Eq.~(\ref{super2hyper}) and this expression is invariant under cyclic symmetry due to nontrivial properties of  three-point kinematics.

\subsection{Amplitudes with massive vectors and tensors}

In this section, we study double-copy amplitudes with massive external fields. Their simplest realization is in a gravitational theory containing massive vectors. We begin by studying amplitudes with gravitons and massive vector fields, and then proceed to include other massless and massive fields. 

Similarly to the massless vectors discussed in the previous section, massive vectors can originate either as the double copies of a massive vector and a massive scalar, or as the double copies of two massive fermions. We start by considering amplitudes between two massive vector fields and a graviton. The first type of massive vectors leads to the double-copy amplitude 
\bea \label{WWh1}
M_3(1W,2W, 3h)   &=&  
- i A_3(1\phi,2\phi, 3A)   A_3(1W,2W, 3A)  \no \\
&& \hskip -1cm  = - i  ( \varepsilon_3 \cdot p_1)  \Big( 2 ( \bep_1 \cdot p_2) ( \bep_2 \cdot \varepsilon_3) + ( \bep_1 \cdot \bep_2) ( \varepsilon_3 \cdot p_1) - (1 \leftrightarrow 2)\Big) .\qquad 
\eea
The second type of double copy leads to 
\bea \label{WWh2}
M_3(1W,2W, 3h)   &=& - i A_3(1\chi,2 \chi, 3A)  A_3(1\tilde \chi,2 \tilde \chi, 3A)  \nn \\
&& \hskip -1cm = i \big( \abra{1}\ket{3}  \ \bra{\qref{3}}\sket{2} 
- (3 \leftrightarrow q_3)  \big) 
\big( \sbra{1}\ket{3}  \ \bra{\qref{3}}\aket{2} 
- (3 \leftrightarrow q_3)  \big)  \ .
\eea
Upon explicit evaluation, one can check that the two expressions are equivalent. This is a consequence of the graviton interacting universally with matter, and an important check that we obtained the correct amplitudes. A change of normalization in the double-copy map would result in these two amplitudes becoming different, i.e. this is a check of the correctness of the normalization of the double-copy fields.

At the level of amplitudes, we can write six gauge-invariant structures with the right little-group indices,\footnote{As in the previous section, we take the two masses to obey $m_1=-m_2=m$.}
\begin{eqnarray}
&& \Big\{ \abra{1}\sket{2} \, \abra{2}\sket{1}  \, (p_1 \cdot \varepsilon_3) , \ \  % \\ &&
\abra{1}\aket{2} \, \sbra{2}\sket{1}  \, (p_1 \cdot \varepsilon_3), \ \
\abra{1} | p_2 |\sket{1} \, \abra{2} | \varepsilon_3 | \sket{2}  - (1 \leftrightarrow 2)  , \ \
\no \\ &&
\qquad \qquad
m_1 \abra{1}\aket{2} \,  \sbra{2} | \varepsilon_3 | \sket{1} , \ 
m_1 \sbra{1}\sket{2} \,  \abra{2} | \varepsilon_3 | \aket{1} , \
m_1 \sbra{1}\aket{2} \,  \sbra{2} | \varepsilon_3 | \aket{1} 
- (1 \leftrightarrow 2) 
\Big\} \ . \quad
\label{sixstructures}
\end{eqnarray}
These structures are however not independent, and it is possible to verify the relations 
\begin{eqnarray}
&& \hskip -1cm 2\big(   \abra{1}\aket{2} \, \sbra{2}\sket{1} - \abra{1}\sket{2} \, \abra{2}\sket{1} \big) (p_1 \cdot \varepsilon_3) +
{1 \over 2} \Big(\abra{1} | p_2 |\sket{1} \, \abra{2} | \varepsilon_3 | \sket{2}  - (1 \leftrightarrow 2)  \Big) \no \\
&& \hskip 5cm 
+ m_1 \abra{1}\aket{2} \,  \sbra{2} | \varepsilon_3 | \sket{1} -
m_1 \sbra{1}\sket{2} \,  \abra{2} | \varepsilon_3 | \aket{1} = 0  
\end{eqnarray}
and
\begin{eqnarray}
&&  \abra{1}\aket{2} \,  \sbra{2} | \varepsilon_3 | \sket{1} +
\sbra{1}\sket{2} \,  \abra{2} | \varepsilon_3 | \aket{1} -
 \sbra{1}\aket{2} \,  \sbra{2} | \varepsilon_3 | \aket{1} -
 \abra{1}\sket{2} \, \abra{2} | \varepsilon_3 | \sket{1}  = 0 \ . \qquad 
\end{eqnarray}
We will use these relations in the remainder of this section and organize 
the three-point matrix elements from the supergravity Lagrangian and from the 
double copy in a basis of gauge-invariant structures 
drawn from Eq.~\eqref{sixstructures}.

In the supergravity Lagrangian we can write three terms that are gauge invariant with respect to the massless vector. The first is a bilinear in the covariant derivative of the massive vector fields,
\begin{equation}
{\cal O}_1=  {D_{[\mu} \overline{W}_{\nu]}} D^\mu W^\nu \ .
\end{equation}
The corresponding three-point matrix element is 
\bea
g_s t^a \langle \overline W  W A \rangle \Big|_{|D W|^2} ~&=&~i g_s t^a ( \bep_1 \cdot  \bep_2 \, (p_1-p_2) \cdot  \varepsilon_3 +  \bep_2  \cdot \varepsilon_3 \, p_2\cdot  \bep_1 -\bep_1  \cdot \varepsilon_3 \, p_1\cdot \bep_2 ) \no \\
&&  \hskip -3cm
= i {g_s t^a \over 2m^2} \left\{ \big(    \abra{1}\aket{2} \, \sbra{2}\sket{1} - \abra{2}\sket{1} \, \abra{1}\sket{2} \, \big) (p_1 \cdot \varepsilon_3) +
 {1 \over 4} \Big(\abra{1} | p_2 |\sket{1} \, \abra{2} | \varepsilon_3 | \sket{2}  - (1 \leftrightarrow 2)  \Big) \right\} ,\no \\
\eea
where $t^a$ is the representation matrix of the massive vectors with respect to the unbroken gauge group and $g_s$ is the coupling constant of the supergravity gauge interactions. Representation indices for the massive vectors are not explicitly displayed.
The second gauge-invariant term in the supergravity Lagrangian has the form 
\be
{\cal O}_2=  \ F^a_{\mu \nu}\overline{W}^\mu t^a W^\nu \ .
\ee
Its contribution to the three-point amplitude is 
\bea
 g_s t^a \langle \overline W W A \rangle \Big|_{W \cdot F \cdot \overline{W} } ~&=& i g_s \, t^a (\bep_1 \cdot \varepsilon_3 \, p_3 \cdot \bep_2 -\bep_2 \cdot \varepsilon_3 \, p_3 \cdot  \bep_1  )  \nn \\
 &=& i {g_s t^a \over 8m^2}  \Big(\abra{1} | p_2 |\sket{1} \, \abra{2} | \varepsilon_3 | \sket{2}  - (1 \leftrightarrow 2)  \Big)   \ .
\eea
Note that in the Yang-Mills Lagrangian with spontaneously-broken gauge symmetry both ${\cal O}_1$ and ${\cal O}_2$ are present with the same normalization, and
\be
\langle \overline W  W A \rangle \Big|_{\rm YM} =\langle \overline W  W A \rangle \Big|_{|D W|^2} + \langle \overline W  W A \rangle \Big|_{W \cdot F \cdot \overline{W} }  \ .
\ee
Lastly, as we have seen in our analysis of the massless theories, an additional operator appears in the supergravity Lagrangian,\footnote{A fourth possible term in the supergravity Lagrangian has the form $D_{(\mu} \overline W_{\nu)} D^\mu W^\nu$ and will not be considered in this section. Together with the three we discuss they exhaust the basis of independent three-point gauge invariants.}
\begin{equation}
 {\cal O}_3=   D\overline W  \wedge A  \wedge DW \ .  
 \end{equation}
The corresponding contribution to the three-vector amplitude is
\be
 t^a \langle  \overline W  W A \rangle \Big|_{ D\overline{W} \wedge A \wedge DW} =
4 i t^a \, \epsilon(\bep_1, \bep_2, \varepsilon_3, p_1, p_2) =
i {t^a } \Big(  \abra{1}\aket{2} \,  \sbra{1} | \varepsilon_3 | \sket{2} +  \sbra{1}\sket{2} \,  \abra{1} | \varepsilon_3 | \aket{2}  \Big)  .
\ee

We now consider amplitudes obtained through the double-copy construction. Specifically, massive fields can have two distinct origins: they can arise as double copies of two massive fermions, or as double copies of one massive vector and one massive scalar. We will take the massive vector either as the vector arising from the double copy of two gluons or as a gluon times an adjoint scalar. Altogether, we have five distinct classes of three-point amplitudes that we need to discuss. We start from the bosonic double copies. The simplest one is\footnote{For amplitudes between fields transforming in matter non-adjoint representations, instead of three-point color-ordered amplitudes, we use three-point amplitudes in which the color factor $T^a_{ij}$ has been stripped.} 
\bea
M_3(1 \overline W,2W, 3A)  &=& -i A_3(1 \overline W,2W, 3A)    A_3(1 \overline \varphi, 2\varphi, 3\phi)\, \no \\
&=& { \lambda \over 2 } \Big( \langle W \overline W A \rangle \Big|_{|D W|^2} + \langle W \overline W A \rangle \Big|_{W \cdot F \cdot \overline{W} } \Big)
\eea
where $\lambda$ is the constant appearing in the three-scalar coupling, which is normalized as $\lambda \int \overline \varphi \phi \varphi$. This leads to the supergravity gauge coupling constant being given by
\begin{equation}
    g_s = {\kappa \lambda \over 4} \ ,
\end{equation}
which is consistent with the result in Ref.~\cite{Chiodaroli:2015wal} (we have restored $\kappa$ in the above equation).  $ A_3(1 \overline W,2W, 3A)$  must be an amplitude from a spontaneously-broken gauge-theory amplitude in order to satisfy color/kinematics duality. 
Another bosonic double copy is \bea
M_3(1 \overline W,2W , 3A)  &=& -i A_3(1 \overline W,2W, 3\phi)  A_3(1\phi, 2\phi, 3A) \nn \\
&=& i { m }  ( \bep_1 \cdot \bep_2) ( \varepsilon_3 \cdot p_1 - \varepsilon_3 \cdot p_2)  \nn \\
&=&    m  \langle W \overline W A \rangle_{|D W|^2} -    m  \langle W \overline W A \rangle_{W \cdot F \cdot \overline{W} } \ .
\eea
In order to have an amplitude of this form, one needs to take $\phi$ to be the Higgs field on the left gauge theory. For this amplitude, the supergravity coupling constant is 
\begin{equation}
 g_s = {\kappa m \over 2} \ .
\end{equation}
There is one last amplitude obtained from a bosonic double copy,
\bea \label{WWA2}
M_3(1 \overline W,2 W, 3A)& =& -i {\epsilon^{ab}\over 4} {\partial \over \partial \auxu_3^{a}}   A_3(1 \overline W,2W, 3A)  {\partial \over \partial \auxu_3^{b}}   A_3(1\overline \varphi,2\varphi, 3A)  \nn \\
&=&  -  \langle W \overline W A \rangle_{  D\overline{W} \wedge A \wedge DW} \,.
\eea
Interestingly, the double copy does not give any term proportional to the mass, and therefore it is the same as in the massless case given in the previous section. 

We now inspect amplitudes in which the massive vectors come from fermionic double copies. There are two such amplitude. The first one is
\bea
M_3(1 \overline W,2W, 3A)  &=&- i {\epsilon^{ab} \over 4} {\partial \over \partial \auxu_3^{a}} A_3(1  \chi,2  \chi, 3A) {\partial \over \partial \auxu_3^{b}} A_3(1 \tilde \chi,2 \tilde \chi, 3A) \nn \\
&=&{ \sqrt{2}}i m \Big[ (\bep_1 \cdot \varepsilon_3 \, p_3 \cdot \bep_2 -\bep_2 \cdot \varepsilon_3 \, p_3 \cdot  \bep_1  )  \Big] \nn \\
&=&{ \sqrt{2}}i m   \langle W \overline W A \rangle_{W \cdot F \cdot \overline{W} } \ .
\eea
The second amplitude is
\bea
M_3(1\overline W,2W, 3A)  &=&-i  A(1  \chi,2  \chi, 3A) A(1 \tilde \chi,2  \tilde  \chi, 3\phi)  \nn \\
&=&  {i \over 2} \sbra{1}\sket{2} \,  \abra{1} | \varepsilon_3 | \aket{2} 
= { m}  \langle W \overline W A \rangle_{|D W|^2}+  {1 \over 4}\langle W \overline W A \rangle_{D\overline{W} \wedge A \wedge DW }  . \qquad
\eea
The result gives both a term proportional to the $C_{IJK}$ tensor and a term that vanishes in the massless limit. 

In formulating the above double-copy constructions, we have paired massive fermions with opposite little-group chiralities. However, we can also choose to have the double-copy map pair fermions with equal chiralities to give supergravity tensors,
\bea
M_3(1\overline B,2B, 3A)  &=&-i  A(1  \chi,2  \chi, 3A) A(1  \chi,2   \chi, 3\phi)  \nn \\
&=& {i \over 2} \abra{1}\aket{2} \,  \abra{1} | \varepsilon_3 | \aket{2}  \ .
\eea
Several comments on these results are now in order. First, we have seen that, in contrast to amplitudes between two vectors and a graviton, which are universal, massive and massless supergravity vectors admit several different kinds of couplings. In particular, vectors with different double-copy origin give rise to different operators in the supergravity Lagrangian. Some of these couplings disappear in the massless limit. This is an indication that the double-copy theory does not admit an unbroken gauge phase. In fact, these classes of amplitudes arise naturally in the double-copy construction for Yang-Mills-Einstein theories with non-compact gauge groups, which always need to have the gauge group be spontaneously broken to a compact subgroup to preserve unitarity. In some cases, one has the choice on how to pair gauge-theory states in the double-copy map. Different choices can result in massive vectors or massive tensors in the output of the double copy. However, the pairing of gauge-theory states depends on their respective gauge-group representations, so that a supergravity state is associated to a gauge-invariant bilinear with states from the two gauge-theory factors. 

In this section, we have used three-point gauge amplitudes for the input of the double copy without studying in detail the theories from which they are originating. In particular, we have not analyzed the constraints coming from color/kinematics duality, which become important for ensuring the consistency of amplitudes at higher-points. Color/kinematics duality may require a delicate balance between the various gauge-group representations and impose constraints on the parameters of the theory. For example, fermions carrying both little-group chiralities can be present in a given representation, which implies that both tensors and vectors will be generated through the double copy. The analysis of the double-copy construction for theories with non-compact gauge groups is beyond the scope of this article and will be carried out in a separate publication \cite{Chiodaroli:2022}.

\section{Conclusion}

In this paper, we have introduced a 5D version of the spinor-helicity formalism, which provides convenient variables valid for 5D massive and massless kinematics, as well as extended on-shell supersymmetry. We have discussed two main applications: the study of amplitudes involving massless and massive Yang-Mills fields and the formulation of the previously-known double-copy construction for $\cN=2$ Maxwell-Einstein and Yang-Mills-Einstein theories in a purely 5D language.   
 Furthermore, we have elaborated on massive self-dual tensor fields. These appear naturally in some supergravity theories in five dimensions and, as such, these theories can be regarded as an important stepping stone towards understanding the mysterious 6D theories of non-abelian tensor fields. 
 
 The 5D ${\cal N}=2$ Maxwell-Einstein and Yang-Mills-Einstein supergravities, which are uniquely determined by their trilinear couplings, are a prime testing ground for understanding whether all supergravity theories exhibit a double-copy structure~\cite{Bern:2010ue,Bern:2019prr}. 
To systematically explore these research directions, efficient methods that make use of the special properties of 5D kinematics are necessary. 
Earlier discussions appeared in Ref.~\cite{Boels:2012ie} and especially Ref.~\cite{Czech:2011dk}, in which a 5D formalism is obtained by dimensionally-reducing the 6D spinor-helicity formalism of Refs.~\cite{Cheung:2009dc, Dennen:2009vk}. See also Refs. \cite{Albonico:2020mge,Geyer:2020iwz} for a discussion of five-dimensional amplitudes in the context of ambitwistor strings and Refs. \cite{Cachazo:2018hqa,Geyer:2018xgb} for methods based on scattering equations.

In this paper, we constructed manifestly-supersymmetric three-point superamplitudes whose components reproduce the amplitudes determined by the 
standard cubic terms of 5D ${\cal N}=2$ supergravity couplings around Minkowski vacua. 
The form of the superamplitudes and the structure of the superfields, with  origins in 6D massless theories, suggest a possible generalization to amplitudes of non-abelian $(2,0)$ self-dual tensor multiplets. However, our analysis, suggesting the absence of covariant expressions corresponding to the candidate spinor-helicity expressions, are consistent with the negative results obtained in Ref.~\cite{Czech:2011dk}. Furthermore, while the candidate four-point formulas for the non-abelian tensor amplitudes are suggestive, they do not pass standard factorization checks, in agreement with the analysis of Ref.~\cite{Cachazo:2018hqa}.

The superamplitudes for maximally supersymmetric gauge and gravity theories are similar to 
the ones obtained through dimensional reduction from 6D. This is a consequence 
of the fact that theories with $\cN=4$ supersymmetry in five dimensions 
always uplift to higher dimensions.
However, superamplitudes with only half-maximal supersymmetry do not always follow a similar pattern. The expressions constructed in Section~\ref{3pointAmps} are 
distinct from the truncations of the corresponding maximally-supersymmetric superamplitudes, and appear to be native to five dimensions. Indeed, the explicit mass dependence in the denominator implies that they cannot be straightforwardly uplifted to six dimensions. 
This is consistent with the fact that not all $\cN=2$ theories in five dimensions possess a higher-dimensional uplift, in close analogy with e.g. 4D ${\cal N}=2$ theories that do not have a 5D uplift. Concretely, whether the superamplitude in Eq.~(\ref{MESGTsuper}) uplifts to six dimensions depends on the particular form of the $C_{IJK}$ tensors. The simplest example of non-upliftable theories discussed in the literature are those in the so-called generic non-Jordan family (see Ref.~\cite{Gunaydin:1983bi}).

In Section~\ref{sec:Neq4WCentralCharge}, we analyzed the 5D ${\cal N}=4$ supersymmetry algebra with a non-singlet central charge and interpreted the central charge as components of the momentum in 10 dimensions. 
We can expose the full 10D Lorentz symmetry by combining the two products of massive 5D spinors in Eq.~\eqref{twospinorproducts} into a covariant Majorana-Weyl 10D spinor
\be
\Lambda_{\cal A}^I =
\left(\begin{matrix}
	\sigma^{i}_{a \dot \alpha}\lambda^{a \dot \alpha}_{A \dot A}  \\
	\sigma^{j}_{\dot a \alpha} \tilde \lambda^{\dot a \alpha}_{A \dot A}  \\
\end{matrix}\right) \ ,
\ee
where $I=i\oplus j$ is an $SO(8)$ little-group vector index, $i\oplus j$ is its decomposition over the subgroup $SO(4)\times SO(4) \subset SO(8)$, and $\sigma^{i}_{a \dot \alpha}, \sigma^{j}_{\dot a \alpha} $ are the respective $SO(4)$ sigma matrices. Finally, the index ${\cal A} = A \otimes \dot A$ is a 16-component Weyl spinor index.
We thus obtain a spinor-helicity-like formalism in 10D, with manifest $SO(1,4)\times SO(5)$ symmetry, which is different from that of Ref.~\cite{Caron-Huot:2010nes}. It would be desirable to explore this formalism further and explicitly construct the low multiplicity superamplitudes of 10D super-Yang-Mills and type IIA/B supergravity.  

Finally, one of the reasons for developing a systematic 5D formalism, and perhaps its main application, is to streamline the study of 5D supergravities  in the double-copy realization of their amplitudes. In five dimensions, ${\cal N}=2$ Maxwell-Einstein and Yang-Mills-Einstein theories are uniquely specified \cite{Gunaydin:1983bi, Gunaydin:1986fg} by the $C_{IJK}$ tensor which enters the three-vector terms in the Lagrangian. We have shown how to extract these $C_{IJK}$ tensors from the double-copy form of the three-vector amplitudes and recovered it for all the homogeneous supergravities, classified in Ref.~\cite{deWit:1991nm} and first given in a double-copy form in Ref.~\cite{Chiodaroli:2015wal}. 
The formalism developed here can be further applied  to the study of gauged supergravities and Yang-Mills-Einstein theories with non-compact gauge groups. We will return on the latter in forthcoming work~\cite{Chiodaroli:2022}. \\

\bigskip

\section*{Acknowledgments}

We would like to thank Donal O'Connell and Oliver Schlotterer for enlightening discussions related to this work. 
This research is supported in part by the Knut and Alice Wallenberg Foundation under grants KAW 2018.0116 ({\it From Scattering Amplitudes to Gravitational Waves}) and KAW 2018.0162, and the Ragnar S\"{o}derberg Foundation (Swedish Foundations' Starting Grant). The work of MC is also supported by the Swedish Research Council under grant 2019-05283. MG would like to thank the Stanford Institute for Theoretical Physics for its hospitality where part of this work was carried out. RR is supported by the US Department of Energy under Grant No. DE-SC00019066.

\newpage

\appendix

\section{Gamma matrices and reality properties for on-shell spinors \label{AppendixA}}

The 5D gamma matrices with lowered $USp(2,2)$ indices are antisymmetric, $\Omega$-traceless and satisfy the following quadratic relations
\bea
&&\Gmat_{AB}^{(\mu} \Omega^{BC} \Gmat_{CD}^{\nu )}= 2 \eta^{\mu\nu} \Omega_{AD}\,, \nn \\
&&\Gmat_{AB}^\mu \Omega^{BC} \Gmat_{CD}^\nu  \Omega^{DA} = 4 \eta^{\mu\nu}\,, \nn \\
&&\Gmat_{AB}^\mu (\Gmat_\mu)_{CD} = 2 (\Omega_{AC} \Omega_{BD}-  \Omega_{AD} \Omega_{BC}) -\Omega_{AB}\Omega_{CD}\,,
\eea
where the first identity is the Clifford algebra.

Under complex conjugation the gamma matrices transform into themselves up to a similarity transform through time-like matrix $\Gmat^{0,AB}$,
\be
(\Gmat^\mu_{AB})^* \equiv \widetilde \Gmat^{\mu,AB} = \Gmat^{0,AC}\Gmat^\mu_{CD}\Gmat^{0,BD} \ .
\ee
Hence it is the time-like gamma matrix
\be
\Gmat^{0,AB}=\Gmat^{0}_{AB} =-(\Gmat^{0}_{BA})^{-1} = \left(\begin{matrix}
	0~ & ~ i \sigma_2~ \\
	i \sigma_2 ~&~ 0 ~ \\
\end{matrix}\right)
\ee
that effectively lowers and raises $USp(2,2)$ indices under complex conjugation. 

By inspection we see that the massive spinors obey the following reality property
\be \label{conjugation}
(|{\bm p}^a\rangle _A)^* \equiv | \widetilde{{\bm p}}_a\rangle^A = \Gmat^{0,AB} |{\bm p}^b \rangle_B E_{ab} \ ,
\ee 
where $E_{ab}$ is a complex unit-determinant matrix ($SL(2,\mathbb{C})$ matrix) that depends on the momentum, the reference vector and the precise choice of spinors in \eqn{USp4spinors}. 

If we consider the spinor parametrization in \eqn{USp4spinors}, and pick a reference vector $\tilde q_\mu=(1,0,0,1,0)$, then the matrix $E_{ab}$ takes the simple form
\be
E_{ab} = \left(\begin{matrix}
	0~ & ~  -x^{-1}~ \\
	x ~&~ 0 ~ \\
\end{matrix}\right) \ ,
\ee
where $x= p_0 + p_3 - \frac{m^2}{p_0 - p_3}$.

Since it has unit determinant, the general $E_{ab}$ matrix satisfies
\be
E_{ac} \epsilon^{cd} E_{bd}= -\epsilon_{ab} \ ,
\ee
and hence it acts as a complex conjugation on the little-group metric. Indeed, we have that
\be
(\epsilon^{ab})^*= - \epsilon_{ab}\,.
\ee
Note that the $\Gmat^{0,AB}$ matrix satisfies the analogous relation
\be
\Gmat^{0,AC} \Omega_{CD} \Gmat^{0,BD} =  -\Omega^{AB} \ ,
\ee
which similarly implies that $(\Omega_{AB})^* = -\Omega^{AB}$.

We can invert the conjugation in \eqn{conjugation}, then we get
\be
\Gmat_{AB}^0  | \widetilde{\bm p}_b\rangle^B  E^{ba} = -  |{\bm p}^a \rangle_A \ .
\ee
This can be interpreted as the statement that the complex conjugated spinor is equal to the original spinor, up to a similarity transform and a sign flip. This is consistent with the spinors being symplectic-Majorana spinors. 

\newpage

\bibliographystyle{JHEP}
\bibliography{litCGJR2021}

\end{document}